\documentclass[preprint,12pt]{elsarticle}

\usepackage{qsymbols}
\usepackage{prooftree}
\usepackage{graphicx}
\usepackage{amsmath}
\usepackage{fancybox}

\usepackage[usenames,dvipsnames]{color}

\newif\ifcomment

\newcommand{\SKIP}[1]{}

\newcommand{\X}{\mathcal X}
\newcommand{\x}{\X}
\newcommand{\astx}{^*\!\x}
\newcommand{\cx}{\ensuremath{^{\scriptstyle\copyright}\!\mathcal X}}
\newcommand{\dx}{^{\scriptstyle d}\!\x}

\newcommand{\lc}{\lambda\mbox{-calculus}}
\newcommand{\lx}{\lambda{\sf x}}
\newcommand{\lmu}{\bar{\lambda}{\mu\tilde{\mu}}}
\newcommand{\lmm}{\lmu}

\newcommand{\lxr}{\lambda{\sf lxr}}

\newcommand{\lxi}{\lambda{\xi}}

\newcommand{\smallastx}{\ensuremath{\scriptstyle \astx}}
\newcommand{\smallx}{\ensuremath{\scriptstyle \mathcal{X}}}
\newcommand{\reduceastx}{\xrightarrow{\smallastx~}}
\newcommand{\reducex}{\xrightarrow{~\smallx~}}

\newcommand{\reduceastxM}{\xrightarrow{\smallastx~}{\small{\!\raisebox{2pt}{+}}}}

\newcommand{\reducexplus}{\xrightarrow{~\smallx~}{\small{\!\raisebox{1pt}{+}}}}

\newcommand{\redplus}{\rightarrow{\small{\!\raisebox{2pt}{+}}}}
\newcommand{\redstar}{\rightarrow{\small{\!\raisebox{2pt}{*}}}}

\newcommand{\Cnt}[1]{C\{#1\}}

\newcommand{\Cntp}[1]{C'\{#1\}}

\newcommand{\Cnts}[1]{C''\{#1\}}

\newcommand{\stm}[2]{#1\preccurlyeq #2}

\newcommand{\I}{\ensuremath{\mathcal I}} 

\newskip \point
\newcommand{\daggerL}{\raise3 \point \hbox{\rotatebox{-40}{$\dagger$}}}
\newcommand{\daggerR}{\raise0 \point \hbox{\rotatebox{40}{$\dagger$}}}

\setbox161=\hbox{$\cdot$}
\setbox162=\hbox{{\raise-2.5\point\copy161\kern-\wd161\raise2.5\point\copy161\kern.5\point
\raise0\point\copy161~}}
\newcommand{\ThreeDots}{\copy162}
\newcommand{\witness}[3]{#1\,\ThreeDots\,{#2}\,\vdash\,{#3}}

\newcommand{\ra}{\rightarrow}

\newcommand{\lin}{\textsf{wf \ensuremath{\astx}-term}}

\newcommand{\threedots}{\ThreeDots}

\newcommand{\Caps}[2]{\langle #1 . #2\rangle}
\newcommand{\Exp}[4]{\widehat{#1}\,#2\,\widehat{#3}\mathop{.}#4}
\newcommand{\Med}[5]{#1\,\widehat{#2}~[#3]~\widehat{#4}\,#5}
\newcommand{\Cut}[4]{#1\widehat{#2}\,\dagger\,\widehat{#3}#4}
\newcommand{\Imp}{\Med}

\newcommand{\Wl}[2]{#2 \odot #1}
\newcommand{\Wr}[2]{#1 \odot #2}
\newcommand{\Wboth}[3]{#2\odot{#1}\odot #3}

\newcommand{\Wlo}[2]{#2 \circledcirc #1}
\newcommand{\Wro}[2]{#1 \circledcirc #2}
\newcommand{\Cbotho}[3]{#2\!\triangleleft\!\Big(#1\Big)\!\triangleright\! #3}

\newcommand{\ren}[2]{\{#1/#2\}} 
\newcommand{\rename}[3]{#1\{#2/#3\}}

\newcommand{\encodex}[1]{\llceil #1 \rrfloor^{\astx}}
\newcommand{\encodeA}[1]{\llceil #1 \rrfloor^{\astx}}
\newcommand{\encodeB}[1]{\llceil #1 \rrfloor^{\x}}

\newcommand{\posswl}[2]{#2\circledcirc #1}
\newcommand{\posswr}[2]{#1\circledcirc #2}
\newcommand{\posswboth}[3]{#2\circledcirc #1\circledcirc #3}
\newcommand{\pwb}{\posswboth}
\newcommand{\poscl}[2]{#2\triangleleft\!\Big(#1\Big)}
\newcommand{\poscr}[2]{\Big(#1\Big)\!\triangleright #2}
\newcommand{\posscboth}[3]{#2\!\triangleleft\!\Big(#1\Big)\!\triangleright\! #3}
\newcommand{\posscbothsmall}[3]{#2\!\triangleleft\!(#1)\!\triangleright\! #3}

\newcommand{\Cl}[4]{#4 {\scriptstyle <} \,
   \raisebox{-3pt}{\ensuremath{\stackrel{\widehat{#2}}{\scriptstyle \widehat{#3}}}}
   \langle#1]}
\newcommand{\Cr}[4]{ [#1\rangle
  \raisebox{-3pt}{\ensuremath{\stackrel{\widehat{#2}}{\scriptstyle\widehat{#3}}}}
  \,{\scriptstyle >} {#4}}

\newcommand{\Cboth}[7]{#4\, {\scriptstyle <} \,
  \raisebox{-3pt}{\ensuremath{\stackrel{\widehat{#2}}{\scriptstyle \widehat{#3}}}}
  \langle#1\rangle
  \raisebox{-3pt}{\ensuremath{\stackrel{\widehat{#5}}{\scriptstyle\widehat{#6}}}}\, {\scriptstyle >}\, #7}

\newcommand{\Clbindlistfin}[7]{(#6,...,\ #7) {\scriptstyle <} \,
  \raisebox{-3pt}{\ensuremath{\stackrel{(\widehat{#2},..., \widehat{#3})}
  {\scriptstyle (\widehat{#4},...,\widehat{#5})}}}
   \langle#1]}

\newcommand{\Crbindlistfin}[7]{ [#1\rangle
  \raisebox{-3pt}{\ensuremath{\stackrel{(\widehat{#2},...,\widehat{#3})}{\scriptstyle (\widehat{#4},...,\widehat{#5})}}}
  \,{\scriptstyle >} (#6,...,#7)}

\newcommand{\Lcut}[4]{#1\widehat{#2}\,\daggerL\,\widehat{#3}#4}
\newcommand{\Rcut}[4]{#1\widehat{#2}\,\daggerR\,\widehat{#3}#4}

\newcommand{\sstermPL}[7]{(#1\widehat{#2}\dagger\widehat{#3}#4)\widehat{#5}\dagger\widehat{#6}{#7}}
\newcommand{\sstermPR}[7]{#1\widehat{#2}\dagger\widehat{#3}(#4\widehat{#5}\dagger\widehat{#6}{#7})}

\newfont{\cir}{wncyr10 scaled \magstephalf}

\newcommand{\lsubs}[4]{\langle\!\langle\widehat{#1}\widehat{#2}\,\daggerL\,\widehat{#3}#4\rangle\!\rangle}
\newcommand{\rsubs}[4]{\langle\!\langle#1\widehat{#2}\,\daggerR\,\widehat{#3}\widehat{#4}\rangle\!\rangle}
\newcommand{\CNT}[3]{#1^{#2}\{#3\}}
\newcommand{\CNTD}[4]{#1^{#2}\{#3,\,#4\}}

\newcommand{\CBGenP}[1]{\Cboth{#1}{{\mathcal I}^P_1}{{\mathcal I}^P_2}{{\mathcal I}^P}
                    {{\mathcal O}^P_1}{{\mathcal O}^P_2}{{\mathcal O}^P}}
\newcommand{\CBGenQ}[1]{\Cboth{#1}{{\mathcal I}^Q_1}{{\mathcal I}^Q_2}{{\mathcal I}^Q}
                    {{\mathcal O}^Q_1}{{\mathcal O}^Q_2}{{\mathcal O}^Q}}

\newtheorem{thm}{Theorem}
\newtheorem{lem}[thm]{Lemma}
\newtheorem{definition}[thm]{Definition}

\newdefinition{rmk}[thm]{Remark}
\newdefinition{exa}[thm]{Example}
\newdefinition{dfn}[thm]{Definition}

\newproof{pf}{Proof}
\newproof{pot}{Proof of Theorem \ref{thm2}}

\newcommand{\figurelinearterms}{
\begin{figure*}[htb]
  \small
{\fboxsep 3mm \fbox{
  \begin{math}
  \small
\begin{array}[c]{cc}
 \multicolumn{2}{c}{
   \prooftree \justifies \Caps{x}{`a}\enskip \lin
  \endprooftree}
\\[6mm]
  \multicolumn{2}{c}{
    \prooftree P\enskip \lin,\enskip x,`b \in  N(P),\enskip `a\notin   N(P)
  \justifies \Exp{x}{P}{`b}{`a}\enskip \lin
  \endprooftree}
\\[6mm]
  \multicolumn{2}{c}{
    \prooftree P, Q\enskip \lin, \enskip `a\in  N(P),\ x\in
  N(Q),\enskip y \notin N(P,Q), \enskip  N(P)\cap  N(Q)=\emptyset
  \justifies \Med{P}{`a}{y}{x}{Q}\enskip \lin
  \endprooftree}
\\[6mm]
  \multicolumn{2}{c}{\prooftree P, Q\enskip \lin, \enskip `a\in N(P),\ x\in N(Q),
  \enskip  N(P)\cap  N(Q)=\emptyset \justifies \Cut{P}{`a}{x}{Q}\enskip
  \lin \endprooftree}
\\[6mm]
  \prooftree P\enskip \lin,\ x\notin N(P) \justifies \Wl{P}{x}\enskip
  \lin \endprooftree
  &
  \prooftree P\enskip \lin,\ `a\notin N(P) \justifies
  \Wr{P}{`a}\enskip \lin \endprooftree
\\[6mm]
  \prooftree P\enskip \lin,\ x,y \in N(P),\ z\notin N(P) \justifies
  \Cl{P}{x}{y}{z} \enskip \lin \endprooftree
  &
  \prooftree P\enskip \lin,\ `a,`b \in N(P),\ `g\notin N(P)
  \justifies \Cr{P}{`a}{`b}{`g} \enskip \lin \endprooftree
\end{array}
\end{math}}}
  \centering
  \vspace*{2mm}
  \caption{$\astx$ terms}
  \label{fig:linearterms}
\end{figure*}
}
\newcommand{\figurefreenames}{
  \begin{figure}[htb]
 \fboxsep 3mm \fbox{\scriptsize%
   \begin{tabular}{c}
      \prooftree%
     \justifies x`: I(\Caps{x}{`a})
     \endprooftree
     \qquad%
     \prooftree%
     \justifies `a`: O(\Caps{x}{`a})
     \endprooftree \\[20pt]
    \prooftree%
  y `: I(P) \quad y\neq x%
  \justifies y `: I(\Exp{x}{P}{`b}{`a})
  \endprooftree
  \qquad %
  \prooftree%
  \justifies `a`: O(\Exp{x}{P}{`b}{`a})
  \endprooftree
  \qquad %
  \prooftree%
  `g `: O(P) \quad `g\neq `b%
  \justifies `g `: O(\Exp{x}{P}{`b}{`a})
  \endprooftree\\[20pt]
\prooftree
\justifies x`: I(\Med{P}{`a}{x}{y}{Q})
\endprooftree
\qquad
\prooftree
z`: I(P)
\justifies z `:I(\Med{P}{`a}{x}{y}{Q})
\endprooftree
\qquad
\prooftree
z`: I(Q) \quad z \neq y
\justifies z `:I(\Med{P}{`a}{x}{y}{Q})
\endprooftree\\[12pt]
\prooftree
`b`: O(P) \qquad `b\neq `a
\justifies `b `:O(\Med{P}{`a}{x}{y}{Q})
\endprooftree
\qquad
\prooftree
`b`: O(Q)
\justifies `b `:O(\Med{P}{`a}{x}{y}{Q})
\endprooftree\\[20pt]
\prooftree
y`:I(P)
\justifies y`:I(\Cut{P}{`a}{x}{Q})
\endprooftree
\qquad
\prooftree
y`:I(Q) \qquad y\neq x
\justifies y`:I(\Cut{P}{`a}{x}{Q})
\endprooftree
\qquad
\prooftree
`b`:O(P) `b\neq `a
\justifies `b`:O(\Cut{P}{`a}{x}{Q})
\endprooftree
\qquad
\prooftree
`b`:O(Q)
\justifies `b`:O(\Cut{P}{`a}{x}{Q})
\endprooftree\\[20pt]
\prooftree
\justifies x `: I(\Wl{P}{x})
\endprooftree
\qquad
\prooftree
y`:I(P)
\justifies y `: I(\Wl{P}{x})
\endprooftree
\qquad
\prooftree
`a`:O(P)
\justifies`a `: O(\Wl{P}{x})
\endprooftree\\[20pt]
\prooftree
x`:I(P)
\justifies x`: I(\Wr{P}{`a})
\endprooftree
\qquad
\prooftree
\justifies`a `: O(\Wr{P}{`a})
\endprooftree
\qquad
\prooftree
`b`:O(P)
\justifies`b `: O(\Wr{P}{`a})
\endprooftree\\[20pt]
\prooftree
\justifies x `: I(\Cl{P}{x_1}{x_2}{x})
\endprooftree
\qquad
\prooftree
y`:I(P) \quad y \neq x_1 \quad y \neq x_2
\justifies y `: I(\Cl{P}{x_1}{x_2}{x})
\endprooftree
\\[20pt]
\prooftree
\justifies `a `: I(\Cr{P}{`a_1}{`a_2}{`a})
\endprooftree
\qquad
\prooftree
`b`:I(P) \quad `b \neq `a_1 \quad `b \neq `a_2
\justifies `b `: I(\Cr{P}{`a_1}{`a_2}{`a})
\endprooftree
\end{tabular}
}
    \caption{Free names in $\astx$}
    \label{fig:freenames}
  \end{figure}
}
\newcommand{\figuretypesystem}{
\begin{figure}[htb]
{\fboxsep 3mm \fbox{\begin{minipage}{13.5cm}\begin{footnotesize}
\[
\prooftree
\justifies {\Caps{x}{`a}\,\ThreeDots~}{x:} A \vdash  {`a:} A %
\using \emph{(ax)}
\endprooftree
\]
\medskip
\[
\prooftree
{P\,\ThreeDots~}`G \vdash  {`a:} A, `D \qquad \quad {Q\,\ThreeDots~} `G', {y:} B \vdash  `D' %
\justifies {\Med{P}{\alpha}{x}{y}{Q}\,\ThreeDots~}`G, `G', {x:}A \rightarrow  B \vdash
`D,`D'  %
 \using \emph{(L$\rightarrow$)} %
\endprooftree
\qquad
\prooftree {P\,\ThreeDots~} `G, {x:} A \vdash  {`a:} B, `D
\justifies {\Exp{x}{P}{\alpha}{\beta}\,\ThreeDots~}`G \vdash  {`b:} A \rightarrow  B, `D %
\using \emph{(R$\rightarrow$)} %
\endprooftree
\]
\medskip
\[
\prooftree {P\,\ThreeDots~} `G \vdash  {`a:} A, `D \qquad \quad {Q\,\ThreeDots~} `G', {x:} A \vdash  `D'%
\justifies {\Cut{P}{\alpha}{x}{Q}\,\ThreeDots~} `G,`G' \vdash  `D,`D' %
\using \emph{(cut)} \endprooftree
\]
\medskip
\[
\prooftree
{P\,\ThreeDots~} `G \vdash  `D
  \justifies
  {\Wl{P}{x}\,\ThreeDots~} `G, {x:} A \vdash `D
\using \emph{(weak\mbox{-}L)}
\endprooftree
\qquad\qquad
\prooftree
{P\,\ThreeDots~} `G \vdash  `D
  \justifies
 {\Wr{P}{`a}\,\ThreeDots~}  `G \vdash  {`a:} A, `D
\using \emph{(weak\mbox{-}R)}
\endprooftree
\]
 \medskip
\[
\prooftree {P\,\ThreeDots~} `G, {x:} A, {y:} A \vdash  `D
 \justifies
{\Cl{P}{x}{y}{z}\,\ThreeDots~} `G, {z:} A\vdash `D
 \using \emph{(cont\mbox{-}L)}
\endprooftree
\qquad\qquad
 \prooftree {P\,\ThreeDots~} `G \vdash  {`a:} A, ` b: A, `D
 \justifies
{\Cr{P}{`a}{`b}{`g}\,\ThreeDots~} `G \vdash  {`g:} A, `D
 \using \emph{(cont\mbox{-}R)}
\endprooftree
\]
\end{footnotesize}\end{minipage}}}
\caption{$\astx$ type system}
\label{fig:TypeSystem}
\end{figure}}

\newcommand{\figuresequentsystemone}{
\begin{figure}[htb]
\begin{small}
\[\prooftree
\justifies A \vdash  A \using
\emph{(ax)} \endprooftree
\]
\medskip
\[
\prooftree
`G \vdash   A, `D \qquad \quad `G',  B \vdash  `D' %
\justifies `G, `G', A \rightarrow  B \vdash
`D,`D'  %
 \using \emph{(L$\rightarrow $)} %
\endprooftree
\qquad\qquad
\prooftree `G,  A \vdash   B, `D
\justifies `G \vdash   A \rightarrow  B, `D %
\using \emph{(R$\rightarrow $)} %
\endprooftree
\]
\medskip
\[
\prooftree `G \vdash   A, `D \qquad \quad `G',  A \vdash  `D'%
\justifies `G,`G' \vdash  `D,`D' %
\using \emph{(cut)} \endprooftree
\]
\medskip
\[
\prooftree
`G \vdash  `D
  \justifies
  `G,  A \vdash `D
\using \emph{(weak\mbox{-}L)}
\endprooftree
\qquad\quad
\prooftree
`G \vdash  `D
  \justifies
 `G \vdash   A, `D
\using \emph{(weak\mbox{-}R)}
\endprooftree
\]
 \medskip
\[
\prooftree `G, A,  A \vdash  `D
 \justifies
`G, A\vdash `D
 \using \emph{(cont\mbox{-}L)}
\endprooftree
\qquad\quad
 \prooftree `G \vdash   A, A, `D
 \justifies
`G \vdash  A, `D
 \using \emph{(cont\mbox{-}R)}
\endprooftree
\]
\medskip
\end{small}
\caption{Sequent system $G1$}
\label{fig:sequentsystemone}
\end{figure}}

\newcommand{\figuresequentsystemthree}{
\begin{figure}[htb]
\begin{small}
\[\prooftree
\justifies \Gamma, A \vdash  A,`D \using
\emph{(ax)} \endprooftree
\]
\medskip
\[
\prooftree
`G \vdash   A, `D \qquad \quad `G,  B \vdash  `D %
\justifies `G, A \rightarrow  B \vdash
`D  %
 \using \emph{(L$\rightarrow $)} %
\endprooftree
\qquad\qquad
\prooftree `G,  A \vdash   B, `D
\justifies `G \vdash   A \rightarrow  B, `D %
\using \emph{(R$\rightarrow $)} %
\endprooftree
\]
\medskip
\[
\prooftree `G \vdash   A, `D \qquad \quad `G,  A \vdash  `D%
\justifies `G \vdash  `D
\using \emph{(cut)} \endprooftree
\]
\end{small}
\caption{Sequent system $G3$}
\label{fig:sequentsystemthree}
\end{figure}}

\newcommand{\figureDlogicalrules}{
\begin{figure}[htb]
\begin{center}{\fboxsep 2mm \fbox{\small
  \begin{math}
  \setlength{\arraycolsep}{2mm}
  \begin{array}{lclcl}
    (ren\mbox{-}L) &:&  \Cut{\Caps{y}{`a}}{`a}{x}{Q}
              &\rightarrow&      \rename{Q}{y}{x}            \\[2mm]
    (ren\mbox{-}R) &:&  \Cut{P}{`a}{x}{\Caps{x}{`b}}
              &\rightarrow&     \rename{P}{`b}{`a}  \\[2mm]
    (ei\mbox{-}insert) &:&
              \Cut{(\Exp{y}{P}{\beta}{`a})}{`a}{x}{(\Med{Q}{\gamma}{x}{z}{R})}
              &\rightarrow&  either \left\{\begin{array}{rr}
                      \sstermPL{Q}{\gamma}{y}{P}{\beta}{z}{R} \\
                      \sstermPR{Q}{\gamma}{y}{P}{\beta}{z}{R}
                      \end{array}
                      \right.
  \end{array}
  \end{math}}}
\end{center}\vspace*{-3mm}
\caption{Logical actions}
\label{fig:DLogicalRules}
\end{figure}}
\newcommand{\figurestructuralactions}{
  \begin{figure*}[htb]
    \small
     \begin{center}\fbox{$
       \setlength{\arraycolsep}{2mm}
       \begin{array}{lclcc}
    \multicolumn{2}{l}{\mbox{\underline{\textit{Left}}}:} && \\[2mm]
    (\daggerL\mbox{-}eras)       &:&
        \Lcut{(\Wr{P}{`a})}{`a}{x}{Q}                                &\rightarrow&
        \Wboth{P}{{\cal I}^{Q}}{{\cal O}^{Q}}  \\
(\daggerL\mbox{-}dupl)      &:&
        \Lcut{(\Cr{P}{`a_1}{`a_2}{`a})}{`a}{x}{Q}                    &\rightarrow&
        P\lsubs{`a_1}{`a_2}{x}{Q}\\[8mm]
    \multicolumn{2}{l}{\mbox{\underline{\textit{Right}}}:} && \\[2mm]
    (\daggerR\mbox{-}eras)   &:&
        \Rcut{P}{`a}{x}{(\Wl{Q}{x})}                         &\rightarrow&
        \Wboth {Q}{{\cal I}^{P}}{{\cal O}^P} \\
(\daggerR\mbox{-}dupl)    &:&
        \Rcut{P}{`a}{x}{(\Cl{Q}{x_1}{x_2}{x})}               &\rightarrow&
        \rsubs{P}{`a}{x_1}{x_2} Q
      \end{array}$}
      \end{center}\vspace*{-3mm}
    \caption{Structural actions}
    \label{fig:structuralactions}
  \end{figure*}}

\newcommand{\figuredeact}{
 \setlength{\arraycolsep}{1mm}
 \begin{figure}[htb]
    \small
    \begin{center}\fbox{$
      \begin{array}{lclcl}
    \mbox{\underline{\textit{Left}}}: && && \\[2mm]
    (\daggerL\mbox{-}deact)   &:&
        \Lcut{P}{\alpha}{x}{Q}   &\rightarrow&
        \Cut{P}{`a}{x}{Q},\enskip\mbox{if}~`a~\mbox{is L-principal for}~P \\[4mm]
    \mbox{\underline{\textit{Right}}}: && && \\[2mm]
    (\daggerR\mbox{-}deact)   &:&
        \Rcut{P}{\alpha}{x}{Q}   &\rightarrow&
        \Cut{P}{`a}{x}{Q},\enskip\mbox{if}~x~\mbox{is L-principal for}~Q
     \end{array}$}
    \end{center}\vspace*{-3mm}
    \caption{Deactivation rules}
    \label{fig:deact}
  \end{figure}}

\newcommand{\figurerightpropagation}{
  \begin{figure*}[htb]
   \small
  \begin{center}{\fboxsep 2mm\fbox{$
       \begin{array}{lclcl}
          (\daggerR exp\mbox{-}prop)    &:&
           \Rcut{P}{`a}{x}{(\Exp{y}{Q}{`b}{`g})}                    &\rightarrow&
           \Exp{y}{(\Rcut{P}{`a}{x}{Q})}{`b}{`g}                                \\[6mm]
          (\daggerR imp\mbox{-}prop_1)    &:&
           \Rcut{P}{`a}{x}{(\Med{Q}{`b}{y}{z}{R})}                 &\rightarrow&
           \Med{(\Rcut{P}{`a}{x}{Q})}{`b}{y}{z}{R},\hspace*{4mm} x\in I(Q)    \\[2mm]
          (\daggerR imp\mbox{-}prop_2)    &:&
           \Rcut{P}{`a}{x}{(\Med{Q}{`b}{y}{z}{R})}                 &\rightarrow&
           \Med{Q}{`b}{y}{z}{(\Rcut{P}{`a}{x}{R})},\hspace*{4mm} x\in I(R)    \\[6mm]
          (\daggerR cut(c)\mbox{-}prop)    &:&
           \Rcut{P}{`a}{x}{(\Cut{\Caps{x}{`b}}{`b}{y}{R})}           &\rightarrow&
           \Cut{P}{`a}{y}{R}                                                   \\[2mm]
          (\daggerR cut\mbox{-}prop_1)    &:&
           \Rcut{P}{`a}{x}{(\Cut{Q}{`b}{y}{R})}             &\rightarrow&
           \Cut{(\Rcut{P}{`a}{x}{Q})}{`b}{y}{R},\hspace*{4mm} x\in I(Q),\ Q\neq\Caps{x}{`b}    \\[2mm]
          (\daggerR cut\mbox{-}prop_2)    &:&
           \Rcut{P}{`a}{x}{(\Cut{Q}{`b}{y}{R})}             &\rightarrow&
           \Cut{Q}{`b}{y}{(\Rcut{P}{`a}{x}{R})},\hspace*{4mm} x\in I(R),\ Q\neq\Caps{x}{`b}    \\[6mm]
          (\daggerR L\mbox{-}eras\mbox{-}prop)    &:&
           \Rcut{P}{`a}{x}{(\Wl{Q}{y})}                     &\rightarrow&
           \Wl{(\Rcut{P}{`a}{x}{Q})}{y},\hspace*{4mm} x\neq y                     \\[2mm]
          (\daggerR R\mbox{-}eras\mbox{-}prop)    &:&
           \Rcut{P}{`a}{x}{(\Wr{Q}{`b})}                    &\rightarrow&
           \Wr{(\Rcut{P}{`a}{x}{Q})}{`b}                                          \\[6mm]
          (\daggerR L\mbox{-}dupl\mbox{-}prop)    &:&
           \Rcut{P}{`a}{x}{(\Cl{Q}{y_1}{y_2}{y})}           &\rightarrow&
           \Cl{\Rcut{P}{`a}{x}{Q}}{y_1}{y_2}{y},\hspace*{4mm} x\neq y           \\[2mm]
          (\daggerR R\mbox{-}dupl\mbox{-}prop)    &:&
           \Rcut{P}{`a}{x}{(\Cr{Q}{`b_1}{`b_2}{`b})}        &\rightarrow&
           \Cr{\Rcut{P}{`a}{x}{Q}}{`b_1}{`b_2}{`b}
       \end{array}$}}
     \end{center}
    \vspace*{-3mm}
    \caption{Right propagation}
    \label{fig:RightPropagation}
\end{figure*}
}
\newcommand{\figureleftpropagation}{
   \begin{figure*}[htb]
    \small
      \begin{center}{\fboxsep 2mm \fbox{$
        \setlength\arraycolsep{0.3em}
          \begin{array}{lclcl}
             (exp\daggerL\mbox{-}prop)   &:&
               \Lcut{(\Exp{x}{P}{\gamma}{`a})}{\beta}{y}{R}                 &\rightarrow&
               \Exp{x}{(\Lcut{P}{`b}{y}{R})}{\gamma}{`a}, \hspace*{4mm} `a\neq `b    \\[6mm]
             (imp\daggerL\mbox{-}prop_1)  &:&
               \Lcut{(\Med{P}{`a}{x}{z}{Q})}{`b}{y}{R}                      &\rightarrow&
               \Med{(\Lcut{P}{`b}{y}{R})}{`a}{x}{z}{Q}, \hspace*{4mm} `b\in O(P)      \\[2mm]
             (imp\daggerL\mbox{-}prop_2)  &:&
               \Lcut{(\Med{P}{`a}{x}{z}{Q})}{`b}{y}{R}                      &\rightarrow&
               \Med{P}{`a}{x}{z}{(\Lcut{Q}{`b}{y}{R})}, \hspace*{4mm} `b\in O(Q)      \\[6mm]
             (cut(c)\daggerL \mbox{-}prop)   &:&
               \Lcut{(\Cut{P}{`a}{x}{\Caps{x}{\beta}})}{\beta}{y}{R}        &\rightarrow&
               \Cut{P}{`a}{y}{R}                                       \\[2mm]
             (cut\daggerL \mbox{-}prop_1)    &:&
               \Lcut{(\Cut{P}{`a}{x}{Q})}{\beta}{y}{R}                      &\rightarrow&
               \Cut{(\Lcut{P}{\beta}{y}{R})}{`a}{x}{Q},\hspace*{4mm} `b\in O(P),\ Q\neq\Caps{x}{`b} \\[2mm]
             (cut\daggerL \mbox{-}prop_2)    &:&
               \Lcut{(\Cut{P}{`a}{x}{Q})}{\beta}{y}{R}                      &\rightarrow&
               \Cut{P}{`a}{x}{(\Lcut{Q}{`b}{y}{R})},\hspace*{4mm} `b\in O(Q),\ Q\neq\Caps{x}{`b} \\[6mm]
             (L\mbox{-}eras\daggerL\mbox{-}prop)     &:&
               \Lcut{(\Wl{P}{x})}{`b}{y}{R}                                 &\rightarrow&
               \Wl{(\Lcut{P}{`b}{y}{R})}{x}                                   \\[2mm]
             (R\mbox{-}eras\daggerL\mbox{-}prop)     &:&
               \Lcut{(\Wr{P}{`a})}{`b}{y}{R}                                &\rightarrow&
               \Wr{(\Lcut{P}{`b}{y}{R})}{`a}, \hspace*{4mm} `a \neq `b        \\[6mm]
             (L\mbox{-}dupl\daggerL\mbox{-}prop)    &:&
               \Lcut{(\Cl{P}{x_1}{x_2}{x})}{`b}{y}{R}                       &\rightarrow&
               \Cl{\Lcut{P}{`b}{y}{R}}{x_1}{x_2}{x}                         \\[2mm]
             (R\mbox{-}dupl\daggerL\mbox{-}prop)    &:&
               \Lcut{(\Cr{P}{`a_1}{`a_2}{`a})}{`b}{y}{R} &\rightarrow&
               \Cr{\Lcut{P}{`b}{y}{R}}{`a_1}{`a_2}{`a}, \hspace*{4mm}
               `a \neq `b
           \end{array}$}}
    \end{center}
      \vspace*{-3mm}
    \caption{Left propagation}
      \label{fig:LeftPropagation}
   \end{figure*}}

\newcommand{\figureencodingx}{
\begin{figure}[htb]
    \setlength\arraycolsep{0.3em}
    \fbox{\begin{math}
           \begin{array}{lcl}
              \encodex{\Caps{x}{`a}}          &:=&
              \Caps{x}{`a}                 \\[5mm]
          \encodex{\Exp{x}{P}{`b}{`a}}    &:=&
                \poscr{\Exp{x}{(\posswboth{\encodex{P}}{x}{\beta})}{`b}{`a}}{`a},\\[5mm]
          \encodex{\Med{P}{`a}{x}{y}{Q}}  &:=&
        \posscboth{\Imp{(\posswr{\encodex{P}}{`a})}{`a}{x}{y}{(\posswl{\encodex{Q}}{y})}}{\I}{\mathcal{O}},\\[2mm]
                && \hspace*{7mm} \mbox{for~} x\notin N(P), x\notin N(Q)\\[5mm]
          \encodex{\Med{P}{`a}{x}{y}{Q}}  &:=&
        \posscboth{\Cl{\Imp{(\posswr{\encodex{P\{x_1/x\}}}{`a})}{`a}{x_2}{y}{(\posswl{\encodex{Q}}{y})}}{x_1}{x_2}{x}}
            {\I}{\mathcal{O}},\\[2mm]
                && \hspace*{7mm} \mbox{for~} x\in N(P), x\notin N(Q)\\[5mm]
          \encodex{\Med{P}{`a}{x}{y}{Q}}  &:=&
        \posscboth{\Cl{\Imp{(\posswr{\encodex{P}}{`a})}{`a}{x_1}{y}{(\posswl{\encodex{Q\{x_2/x\}}}{y})}}{x_1}{x_2}{x}}
            {\I}{\mathcal{O}},\\[2mm]
                && \hspace*{7mm} \mbox{for~} x\notin N(P), x\in N(Q)\\[5mm]
          \encodex{\Med{P}{`a}{x}{y}{Q}}  &:=&
        \posscboth{\Cl{\Cl{\Imp{(\posswr{\encodex{P\{x_1/x\}}}{`a})}{`a}{x_2}{y}{(\posswl{\encodex{Q\{x_3/x\}}}{y})}}{x_1}{x_2}{t}}{t}{x_3}{x}}
            {\I}{\mathcal{O}},\\[2mm]
                && \hspace*{7mm} \mbox{for~} x\in N(P), x\in N(Q)\\[5mm]
          \encodex{\Cut{P}{`a}{x}{Q}}    &:=&
        \posscboth{\Cut{(\posswr{\encodex{P}}{`a})}{`a}{x}{(\posswl{\encodex{Q}}{x})}}{\I}{\mathcal{O}},\\[2mm]
    \end{array}
    \end{math}}
    \caption{Encoding the $\X$-terms into $\astx$}
    \label{fig:encodingx}
\end{figure}}

\newcommand{\figurexsyntax}{
\begin{figure}[htb]
  \begin{center}
  \doublebox{
  $
  \begin{array}{lcll}
P,Q    &::=&       \Caps{x}{`a}           &    \quad \textit{capsule}   \\[1mm]
       &\mid&      \Exp{x}{P}{`b}{`a}     &    \quad \textit{exporter}    \\[1mm]
       &\mid&      \Med{P}{`a}{x}{y}{Q}   &    \quad \textit{importer}  \\[1mm]
       &\mid&      \Cut{P}{`a}{x}{Q}      &    \quad \textit{cut}
  \end{array}
  $}\end{center}
\vspace*{-4mm}
\caption{The syntax of $\x$}
\label{fig:xsyntax}
\end{figure}}

\newcommand{\figurexactivation}{
\begin{figure}[ht]
\begin{center}\fbox{$
\begin{small}
\setlength\arraycolsep{2mm}
 \begin{array}{rcl}
     (act\mbox{-}L) &:& \Cut{P}{`a}{x}{Q} \ \rightarrow \Lcut{P}{`a}{x}{Q},\enskip
          \mbox{if $`a$ not freshly introduced by $P$} \\[2mm]
     (act\mbox{-}R) &:& \Cut{P}{`a}{x}{Q}\ \rightarrow\Rcut{P}{`a}{x}{Q},\enskip
          \mbox{if $x$ not freshly introduced by $Q$}
 \end{array}
 \end{small}
 $}\end{center}\vspace*{-3mm}
\caption{Activation rules in $\x$}
\label{fig:xactivation}
\end{figure}}

\newcommand{\figurexlogical}{
\begin{figure}[htb]
\begin{center}\fbox{
  \begin{math}
  \setlength{\arraycolsep}{2mm}
  \begin{small}
  \begin{array}{lclcl}
        (ren\mbox{-}R) &:&  \Cut{P}{\alpha}{x}{\Caps{x}{`b}}
            ~~~\rightarrow~~~   \rename{P}{`b}{`a}\\[3mm]
        (ren\mbox{-}L) &:&  \Cut{\Caps{y}{`a}}{\alpha}{x}{Q}
                ~~~\rightarrow~~~
        \rename{Q}{y}{x}\\[3mm]
              \Cut{(\Exp{y}{P}{\beta}{`a})}{`a}{x}{(\Med{Q}{\gamma}{x}{z}{R})}
              &\rightarrow&  either\enskip\left\{\begin{array}{rr}
                      \sstermPL{Q}{\gamma}{y}{P}{\beta}{z}{R} \\
                      \sstermPR{Q}{\gamma}{y}{P}{\beta}{z}{R}
                      \end{array}
                      \right. \\[1mm]
    &&\multicolumn{3}{l}{`a\notin N(P),~x\notin N(Q),~x\notin N(R)}
  \end{array}
  \end{small}
  \end{math}}
\end{center}\vspace*{-3mm}
\caption{Logical rules in $\x$}
\label{fig:xlogical}
\end{figure}}

\newcommand{\figurexmixL}{
   \begin{figure}[htb]
    \small
      \begin{center}\fbox{$
        \setlength\arraycolsep{0.3em}
          \begin{array}{lcrcl}
             (\daggerL\mbox{-}eras)   &:&
               \Lcut{\Caps{x}{`a}}{`b}{y}{R}                 &\rightarrow&
               \Caps{x}{`a}, \hspace*{4mm} `a\neq `b    \\[2mm]
             (\daggerL\mbox{-}deact)   &:&
               \Lcut{\Caps{x}{`b}}{`b}{y}{R}                 &\rightarrow&
               \Cut{\Caps{x}{`b}}{`b}{y}{R}\\[2mm]
             (\daggerL\mbox{-}prop)   &:&
               \Lcut{(\Exp{x}{P}{\gamma}{`a})}{\beta}{y}{R}                 &\rightarrow&
               \Exp{x}{(\Lcut{P}{`b}{y}{R})}{\gamma}{`a}, \hspace*{4mm} `a\neq `b    \\[2mm]
             (\daggerL\mbox{-}prop\mbox{-}dupl\mbox{-}deact)   &:&
               \Lcut{(\Exp{x}{P}{\gamma}{`b})}{\beta}{y}{R}                 &\rightarrow&
               \Cut{(\Exp{x}{(\Lcut{P}{`b}{y}{R})}{\gamma}{`b})}{`b}{y}{R}   \\[2mm]
             (\daggerL\mbox{-}prop\mbox{-}dupl_1)  &:&
               \Lcut{(\Med{P}{`a}{x}{z}{Q})}{`b}{y}{R}                      &\rightarrow&
               \Med{(\Lcut{P}{`b}{y}{R})}{`a}{x}{z}{(\Lcut{Q}{`b}{y}{R})}\\[2mm]
             (\daggerL\mbox{-}(c)\mbox{-}prop\mbox{-}deact)   &:&
               \Lcut{(\Cut{P}{`a}{x}{\Caps{x}{\beta}})}{\beta}{y}{R}        &\rightarrow&
               \Cut{(\Lcut{P}{`b}{y}{R})}{`a}{y}{R}                                       \\[2mm]
             (\daggerL\mbox{-}prop\mbox{-}dupl_2)    &:&
               \Lcut{(\Cut{P}{`a}{x}{Q})}{\beta}{y}{R}                      &\rightarrow&
               \Cut{(\Lcut{P}{\beta}{y}{R})}{`a}{x}{(\Lcut{Q}{\beta}{y}{R})},\hspace*{4mm}
               Q\neq\Caps{x}{`b}\\[2mm]
             (\daggerL\mbox{-}gc)   &:& \Lcut{P}{\alpha}{x}{Q}   &\rightarrow&
                P,\enskip\mbox{if}~`a \notin N(P)
           \end{array}$}
    \end{center}
      \vspace*{-3mm}\caption{Left propagation (erasure/duplication/deactivation) in $\x$}
      \label{fig:xmixL}
   \end{figure}}

\newcommand{\figurexmixR}{
   \begin{figure}[htb]
    \small
      \begin{center}\fbox{$
        \setlength\arraycolsep{0.3em}
          \begin{array}{lclcl}
             (\daggerR\mbox{-}eras)   &:&
               \Rcut{P}{`a}{x}{\Caps{y}{`b}}                 &\rightarrow&
               \Caps{y}{`b}, \hspace*{4mm} x\neq y    \\[2mm]
             (\daggerR\mbox{-}deact)   &:&
               \Rcut{P}{`a}{x}{\Caps{x}{`b}}                 &\rightarrow&
               \Cut{P}{`a}{x}{\Caps{x}{`b}}\\[2mm]
             (\daggerR\mbox{-}prop)   &:&
               \Rcut{P}{`a}{x}{(\Exp{y}{Q}{`g}{`b})}                 &\rightarrow&
               \Exp{y}{(\Rcut{P}{`a}{x}{Q})}{`g}{`b}    \\[2mm]
             (\daggerR\mbox{-}prop\mbox{-}dupl_1)  &:&
               \Rcut{P}{`a}{x}{(\Imp{Q}{`b}{y}{z}{R})}                      &\rightarrow&
               \Imp{(\Rcut{P}{`a}{x}{Q})}{`b}{y}{z}{(\Rcut{P}{`a}{x}{R})}, \hspace*{4mm} x\neq y\\[2mm]
             (\daggerR\mbox{-}prop\mbox{-}dupl\mbox{-}deact)  &:&
               \Rcut{P}{`a}{x}{(\Imp{Q}{`b}{x}{z}{R})}                      &\rightarrow&
               \Cut{P}{`a}{x}{(\Imp{(\Rcut{P}{`a}{x}{Q})}{`b}{x}{z}{(\Rcut{P}{`a}{x}{R})})}\\[2mm]
             (\daggerR\mbox{-}(c)\mbox{-}prop\mbox{-}deact)   &:&
               \Rcut{P}{`a}{x}{(\Cut{\Caps{x}{`b}}{`b}{y}{R})}        &\rightarrow&
               \Cut{P}{`a}{y}{(\Rcut{P}{`a}{x}{R})}\\[2mm]
             (\daggerR\mbox{-}prop\mbox{-}dupl_2)    &:&
               \Rcut{P}{`a}{x}{(\Cut{Q}{`b}{y}{R})}                      &\rightarrow&
               \Cut{(\Rcut{P}{`a}{x}{Q})}{`b}{y}{(\Rcut{P}{`a}{x}{R})},\hspace*{4mm}
               Q\neq\Caps{x}{`b}\\[2mm]
             (\daggerR\mbox{-}gc)   &:& \Rcut{P}{\alpha}{x}{Q}   &\rightarrow&
              Q,\enskip\mbox{if}~x\notin N(Q)
           \end{array}$}
    \end{center}
      \vspace*{-3mm}\caption{Right propagation (erasure/duplication/deactivation) in $\x$}
      \label{fig:xmixR}
   \end{figure}}

\newcommand{\figurextypes}{
\begin{figure}[htb]
{\fboxsep 3mm \fbox{\begin{minipage}{13.2cm}
\begin{footnotesize}
\[
\prooftree
\justifies \witness{\Caps{x}{`a}}{`G,x:A}{`a:A,`D} \using
\emph{(axiom)}
\endprooftree
\]
\medskip
\[
\prooftree
\witness{P}{`G}{`a:A,`D}\qquad\witness{Q}{`G,x:B}{`D}
\justifies \witness{\Med{P}{`a}{y}{x}{Q}}{`G, y:A\ra B}{`D} %
\using \emph{($\rightarrow$L)} %
\endprooftree
\qquad
\prooftree
    \witness{P}{`G, x:A}{`a:B, `D} %
    \justifies
    \witness{\Exp{x}{P}{`a}{`b}}{`G}{`b:A\rightarrow B, `D}%
 \using \emph{($\rightarrow$R)} %
\endprooftree
\]
\medskip
\[
\prooftree
    \witness{P}{`G}{`a:A, `D}\qquad \witness{Q}{`G,x:A}{`D}
    \justifies
    \witness{\Cut{P}{`a}{x}{Q}}{`G}{`D}
\using \emph{(cut)}
\endprooftree
\]
\end{footnotesize}
\end{minipage}}}
\caption{$\X$ type system}
\label{fig:xtypes}
\end{figure}}

\newcommand{\figureencodingB}{
\begin{figure}[htb]
    \begin{center}\fbox{
    $\begin{array}{rcl}
    \encodeB{\Caps{x}{\alpha}}  &:=& \Caps{x}{\alpha}\\[2mm]
    \encodeB{\Exp{x}{P}{`b}{`a}}    &:=& \Exp{x}{\encodeB{P}}{`b}{`a}\\[2mm]
    \encodeB{\Imp{P}{`a}{x}{y}{Q}}  &:=& \Imp{\encodeB{P}}{`a}{x}{y}{\encodeB{Q}}\\[2mm]
    \encodeB{\Cut{P}{`a}{x}{Q}}     &:=& \Cut{\encodeB{P}}{`a}{x}{\encodeB{Q}} \\[2mm]
    \encodeB{\Cl{P}{y}{z}{x}}   &:=& \encodeB{P}\ren{x}{y}\ren{x}{z}\\[2mm]
    \encodeB{\Cr{P}{`b}{`g}{`a}}    &:=& \encodeB{P}\ren{`a}{`b}\ren{`a}{`g}\\[2mm]
    \encodeB{\Wl{P}{x}}     &:=& \encodeB{P}\\[2mm]
    \encodeB{\Wr{P}{`a}}        &:=& \encodeB{P}
     \end{array}$}
     \end{center}
\label{fig:encodeB}
\caption{Encoding the $\astx$-terms into $\x$}
\end{figure}}

\journal{Theoretical Computer Science}

\begin{document}

\begin{frontmatter}

\title{Computational interpretation of classical logic\\ with explicit structural rules}

\author[1]{S. Ghilezan} \ead{gsilvia@uns.ac.rs}
\author[2]{P. Lescanne} \ead{pierre.lescanne@ens-lyon.fr}
\author[3]{D. \v Zuni\'c}  \ead{dragisa.zunic@gmail.com}

\address[1]{University of Novi Sad, Faculty of Technical Sciences,  Serbia}
\address[2]{University of Lyon, \' Ecole Normal Sup\' erieure de Lyon, France}
\address[3]{Faculty of Economics and Engineering Management, Novi Sad, Serbia}

 \date{\today}

\begin{abstract}
We present a calculus providing a Curry-Howard
correspondence to classical logic represented in the sequent
calculus with explicit structural rules, namely weakening and
contraction. These structural rules introduce explicit erasure and
duplication of terms, respectively. We present a type system for
which we prove the type-preservation under reduction. A mutual
relation with classical calculus featuring implicit structural
rules has been studied in detail. From this analysis we derive
strong normalisation property.
\end{abstract}


\end{frontmatter}


\section*{Introduction}
\label{sec:intro}

The fundamental connection between logic and computation, known as
the Curry-Howard correspondence or {\em formulae-as-types, proofs-as-term} and
{\em proofs-as-programs pa\-ra\-digm},  relates logical and computational
systems.



Gentzen's natural deduction is a well established formalism for
expressing proofs. Church's simply typed $\lambda$-calculus is a
core formalism for writing programs. {\em Simply typed} $\lambda$-calculus represents a computational
interpretation of {\em intuitionistic natural deduction}: formulae correspond to types, proofs to terms/programs
and simplifying a proof corresponds to executing a program. In its traditional form, terms in the $\lambda$-calculus encode
proofs in intuitionistic natural deduction; from another perspective the proofs serve
as typing derivations for the terms.
This correspondence was  discovered in the late 1950s and early 1960s independently in logic by Curry, later formulated by Howard; in category theory, Cartesian Closed Categories, by Lambek; and in mechanization of mathematics, the language Automath,   by de Brujin.

Griffin extended the Curry-Howard correspondence to classical logic in his
seminal 1990 paper~\cite{Gri90}, by observing that classical
tautologies suggest typings for certain control operators.
This initiated a vigorous line of
research: on the one hand classical calculi can be seen as pure programming languages
with explicit representations of control, while at the same time terms can be tools for
extracting the constructive content of classical proofs.
The $\lambda \mu$-calculus of Parigot \cite{Par92} expresses the computational content of {\em classical
natural deduction} and has been the basis of a number of
investigations into the relationship between classical logic and theories of
control in programming languages. 

Computational interpretation of {\em sequent-style logical systems} has come into the picture much later, by the end of 1990s. There were several attempts, over the years, to design a term calculus which would embody the Curry-Howard correspondence for {\em intuitionistic sequent
logic}. The first calculus accomplishing this task is Herbelin's $\bar{\lambda}$-calculus \cite{Her95}.
Recent interest in the Curry-Howard correspondence for
intuitinistic sequent logic \cite{Her95,bareghil00,jesTLCA07,espighilivet07}
made it clear that the computational content of sequent
derivations and cut-elimination can be expressed through an
extension of the $\lambda$-calculus.
In the classical setting, there are several term calculi based
on {\em classical sequent logic}, in which terms
unambiguously encode sequent derivations and reduction corresponds to cut
elimination: Barbanera and Berardi's Symmetric Calculus \cite{BarBer94},  Curien-Herbelin's $\overline{\lambda} \mu \widetilde{\mu}$-calculus~\cite{CurHer00}, Urban-Bierman's calculus~\cite{urbabier99}, Wadler's Dual Calculus \cite{wadlerdual}.
In contrast to natural deduction proof systems, sequent calculi exhibit inherent
symmetries in proof structures which create
technical difficulties in analyzing the reduction properties of
these calculi \cite{dougghillesclika05,dougghillesc07,ghil07}.

The tutorial entitled ``Computational interpretations of logics" given by the first author of this paper at ICTAC 2011 in Johannesburg, South Africa, presented a comprehensive overview and a comparison of computational interpretations of
intuitionistic and classical logic both in natural deduction and sequent-style setting. In this paper our focus is on the computational interpretations of classical sequent calculus, with explicit structural rules of weakening and
contraction.

$\astx$ has been designed to provide a correspondence `a l\`a'
Curry-Howard for the standard formulation of classical sequent
calculus, with explicit structural rules (weakening and
contraction). The direct correspondence between proofs and terms
is achieved by using the technique of labeling formulas by names.
These names are used to build terms so that the structure of a
term captures the original structure of a corresponding proof.
Furthermore, the computation of terms is defined in a way that
mirrors the proof-transformation, that is, the cut-elimination.

The inspiration for $\astx$ comes from two sources. On the one
hand, the direct predecessor is the classical term language called
$\x$. On the other hand, a very strong influence comes from the
intuitionistic field and most notably the work on
the~$\lxr$-calculus.
\begin{center}
\fboxsep2mm\fbox{\parbox{3cm}{\centerline{$\x\raisebox{-2cm}{$\includegraphics[width=2cm]{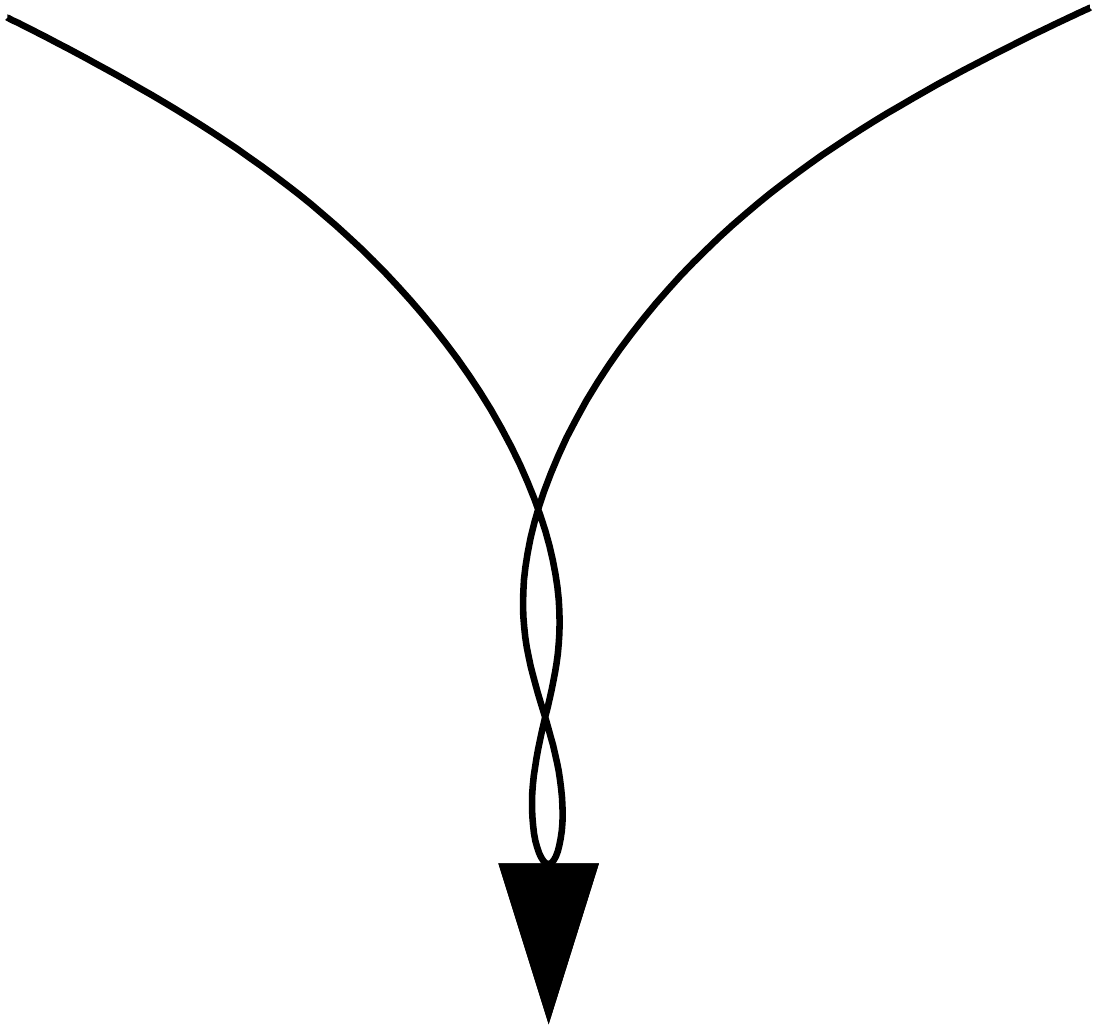}$}\lxr$}\vspace*{2mm}
\centerline{$\astx$~~}}}
\end{center}
\noindent In our study we try to respect the underlying principles
of these works, and implement them in a way that preserves their
good properties.

As a first contribution of this paper, we design $\astx$,
which represents the computational interpretation of classical
sequent logic with explicit structural rules of contraction and
weakening. Further, we propose a simply typed system for which we
prove the witness reduction property. We relate the explicit and
implicit treatment of structural rules by mutual encoding of
$\astx$ and $\x$. Finally, these results leads us to prove strong
normalisation of simply typed $\astx$.

\paragraph{Related work}

The $\x$ calculus is a  term
language, introduced in \cite{vBLL} and studied in more detail in
\cite{vBL}. It is a low level language which can easily encode various
other calculi and which captures the structure of classical proofs represented in the sequent calculus, espcially cut-elimination. Some of its properties are non-determinism, non-confluence and strong normalizationfor typed terms.

Some closely related computational interpretations have been presented earlier. First of them is the so-called local cut-elimination procedure presented in \cite{BieUrb99}. It is one of the three cut-elimination procedures studied in detail in \cite{Urb00}. A term assignment is given for proofs in the  classical sequent calculus (formulated with completely implicit structural rules). Then this term language was used as a tool to show the properties of classical sequent calculus. Most importantly, it enabled the authors to use the term-rewriting techniques in order to prove the strong normalization of cut-elimination in classical logic.

A second computational interpretation, very close to $\x$, has been presented by Lengrand in \cite{LenCBV}, under the name $\lxi$-calculus. There it was studied in relation with $\lmm$-calculus of~\cite{CurHer00}, and it was used to infer the strong normalization for \ensuremath{\lmm}.

Although there are differences these three formulations are very close. The syntaxes of
\ensuremath{\lxi} and \ensuremath{\x} are the same (there are minor differences such as the use of
\ensuremath{\dagger_{\mbox{v}}} in the first, instead of \ensuremath{\daggerL} in the second). Both
the syntax and the reduction rules of \ensuremath{\lxi} are said to be (in \cite{LenCBV}) the
subsystems of Urban's local cut-elimination procedure (\ensuremath{{\mathcal T}^\leftrightarrow},
\ensuremath{\xrightarrow{loc}}) (see \cite{Urb00}). However, some differences in the set of
reductions exist.

\SKIP{
The so-called logical rules are identical in \ensuremath{\lxi,\x}, but not in
Urban's calculus.  Notice that Urban includes a rule for garbage collection, which appears neither
in \ensuremath{\lxi} nor in \ensuremath{\x}. Moreover, there is a rule for cut-propagation over a
capsule (as an exception rule from general propagation philosophy) which appears in Urban's work and
\ensuremath{\lxi}, but not in~\ensuremath{\X}. Finally there is a subtle difference in Urban's
activation rules, namely he has an extra condition which prevents the activation of a cut when the
term has an active cut on the top.
}

Let us recall here the philosophy behind these calculi.  Urban \cite{Urb00} was partly inspired by
Danos et al.~\cite{DJS97} who consider Gentzen's sequent calculus as a programming language. Their
cut-elimination procedure is called \ensuremath{\mbox{LK}^{tq}}. It is strong normalizing, confluent
and strongly connected to linear logic proof nets.  Confluence is obtained by assigning color
annotations to formulas, which restricts cut-reductions so that the critical pair does not
arise. Confluence is essential in \ensuremath{\mbox{LK}^{tq}} because it enabled the authors to
exploit the strong normalization result of proof nets in linear~logic. However Urban reveals all the
details of the complex classical cut-elimination, by developing a term-notation for proofs, whereas
in \ensuremath{\mbox{LK}^{tq}} concepts are presented informally.

Moreover, it has been shown in \cite{Urb00, BieUrb01}, using the results of~\cite{BarBer97}, that not all normal forms are reachable using the \ensuremath{\mbox{LK}^{tq}} interpretation. Secondly, the restrictions introduced by using the colors are not needed to ensure strong normalization.

$\X$ departs from the traditional doctrine of intuitionistic logic, where computation is an equality
preserving operation on proofs. Instead, $\X$ accepts that cut-elimination may or may not preserve proof
equality, and that non-determinism is a natural feature of classical logic.

\SKIP{
Urban's term
language was designed in according to the following principles:

\begin{itemize}
\item Cut-elimination should not restrict the collection of normal forms (more precisely, essential normal forms) that are reachable.
\item  Cut-elimination should be strongly-normalizing.
\item  Cuts should be allowed to pass over other cuts, namely over other inactive cuts.
\end{itemize}
There are however a few restrictions. Namely, active cuts are not allowed to pass over other active cuts, and once the cut is activated it cannot be deactivated at will (the propagation has to continue in the same direction).
}

Although mainly concerned with the computational content of classical logic, the ideas presented in
this paper come partly from intuitionistic logic, primarily from the
$\lxr$-calculus~\cite{KesLen05,KesLen07} which had a significant influence.
The $\lxr$-calculus extends $\lx$ \cite{BlooRose95,RoseBlooLang:jar2011} by operators for erasure and duplication in the
same way as $\astx$ extends~$\x$.  The intuitionistic calculi, $\lx$ and~$\lxr$ are related as are
~$\x$ and~$\astx$.

\SKIP{
Moreover, if we consider the propagation rules of \ensuremath{\x} and
\ensuremath{\astx} to be the analog of explicit substitution, then the calculi which do not have
propagation rules, namely diagrammatic calculus (see Part~\ref{part:diag}) and the \ensuremath{\cx}
(see Part~\ref{part:cPART}) could be related to \ensuremath{`l}-calculus which has an implicit
substitution. This should not be taken strictly because the erasure and the duplication are explicit
in~\ensuremath{\dx} and~\ensuremath{\dx}, which is not the case in the~\ensuremath{`l}-calculus
}

The $\lxr$-calculus was created as an attempt to relate the two elementary decompositions, namely, the decomposition of intuitionistic connectives in linear logic, and the decomposition of a meta-level \ensuremath{\lc} substitution. The meta-substitution can be decomposed into more atomic steps, represented within the language \cite{ACCL91}, thus bringing the theoretical work closer to the actual implementations. It has been shown in \cite{KesLen05} that there exists a very strong relation between \ensuremath{\lxr}-calculus and linear logic proof-nets.

Some works have considered the relation between $\x$ and the $\pi$-calculus. The $`p$-calculus, \cite{Mil95,SanWal01}, is able to describe concurrent computations, including the communication between processes. The configurations of the interacting processes may change during the computation.
The relation of $\X$ and $\pi$-calculus has been recently presented in \cite{fromXtoPi}, where the~\mbox{$\x$~calculus} is encoded into $\pi$. This paper seeks for the intuition to what is computational meaning of cut-elimination from the point of view of $`p$.

Some remarks aiming at essential points related to concurrency were given earlier by Urban \cite{Urb00}. He suggested how a form of weak communication can be implemented, using quantifiers, into the classical sequent calculus. Besides that, it was noted that the approach where reduction is not seen as an equality preserving operation, is a standard approach in the calculi of concurrency.
Moreover, the substitution mechanism in $\x$-like calculi in which only names may participate, is closer to the~ $\pi$-calculus than the substitution mechanism defined in the~$`l$-calculus which involves terms.

\paragraph{Outline of the paper}
Section~\ref{sec:sequents} is a brief overview of the sequent style classical logical systems.
Section~\ref{sec:x} deals with $\x$ calculus: its syntax, reduction rules, types systems and basic properties.
In Section~\ref{sec:astx} we propose the syntax and operational semantics of $\astx$ as well as the simply typed system.
Section~\ref{sec:xastx} provides the relation between the two calculi.
\tableofcontents

\section{Sequent calculi $G1$ and $G3$}
\label{sec:sequents}

The basic Genzen systems for classical and intuitionistic logic
denoted as $G1, G2$ and $G3$ are formalized in \cite{klee} and
later revisited in \cite{SchTro}. In brief, the essential
difference between $G1$ and $G3$ is the presence or absence of
explicit structural rules. The distinguishing point in the case of
$G2$ is the use of the so-called mix instead of a cut rule.
Although here we focus on the classical systems, we remark that
the intuitionistic systems are obtained from classical ones by
restricting sequents to having only one formula in the~succedent.

\paragraph*{The system $G1$}
Among the three systems  presented by Kleene~\cite{klee}, $G1$ is
the closest to Gentzen's original formulation~\cite{gentzen35}. Despite the fact that Gentzen and Kleene
present explicitly  exchange rules, which is not the case here, we
keep the name~$G1$ (Figure~\ref{fig:sequentsystemone}). Latin
symbols $A, B, ...$ are used to denote formulas and Greek symbols
$`G, `D,`G',`D',...$ to denote contexts, which are in this
framework multisets of formulas. Exchange rules are handled by
multisets instead of lists, whereas the other structural
rules, namely \emph{weakening} and \emph{contraction} are
explicitly given. The axiom rules do not involve arbitrary
contexts. Inference rules with two premises, namely $(L \to)$ and
\emph{(cut)}, are given in the context-splitting style, which
means that when looking bottom-up the contexts of a conclusion is
split by premises. It has been shown in \cite{SchTro} that if a
context-sharing style was applied one obtains an equivalent
system, i.~e., a system that proves the same sequents.
\figuresequentsystemone

\paragraph*{The system $G3$} The sequent system~$G3$ is obtained from $G1$ by making all
structural rules parts of the remaining rules with appropriate forms.
In other words, there is no explicit structural rules. Instead
structural rules are hidden in the new presentation of the
logical rules and of the cut-rule, and thus performed automatically.

This system has been mentioned as $G3a$ in \cite{klee} and formalized as classical $G3$ in \cite{SchTro}.
It is presented by Figure \ref{fig:sequentsystemthree}, where $A, B, ...$ range over formulas, while contexts
$\Gamma, \Delta, ...$ are finite \emph{sets} of formulas.
\figuresequentsystemthree

\noindent Inference rules with two premises are given in the context-sharing style. The definition of the
axiom rule involves contexts, thus allowing arbitrary formulas to be introduced at that level, i.e., weakening
rule is hidden in the form of the axiom.

\section{The $\x$ calculus}
\label{sec:x}

This section presents $\x$ which is, together with $\lxr$, a
predecessor of $\astx$. The design of $\astx$ has been directly
inspired by~$\x$.

$\x$ was first presented in van Bakel, Lescanne and Lengrand in \cite{vBLL}. The origin of the
language is in the notations for classical sequent proofs by Urban \cite{Urb00}, introduced as a tool
to express the cut-elimination procedure as a term rewriting system, which later allowed him to prove
strong normalization of cut-elimination. A close variant of the language has been studied by Lengrand in
relation with the~$\lmu$-calculus, in a calculus he called $\lambda\xi$~\cite{LenCBV}.

It is argued in \cite{Urb00} that non-determinism, although it leads to non-confluence,
should be considered as an intrinsic property of classical logic. This point of view was
taken in some earlier works, for example \cite{BarBer94,Her95,BarBer97} and more recently in
\cite{LocSol,Hyl00}. This means that, in classical logic, we depart from the traditional intuitionistic
(and linear) logic doctrine, where cut-elimination is an equality preserving operation on proofs.

\subsection{The syntax}
The $\x$ calculus corresponds to a sequent system with implicit
structural rules (Figure \ref{fig:sequentsystemthree}. Since we
consider only the implicative fragment, the only inference rules
are axiom, cut, left-arrow introduction and right-arrow
introduction. Therefore, in the~$\x$ calculus there are four
constructors (see Figure \ref{fig:xsyntax}). \figurexsyntax

The term \emph{capsule} corresponds to an axiom rule, \emph{cut} corresponds to a cut-rule,
\emph{importer} corresponds to left-arrow introduction rule and \emph{exporter} corresponds to
right-arrow introduction rule.\footnote{In the original papers importer and exporter were called import and mediator.}
The syntax is then extended by two \emph{active cuts} that reflect the non-deterministic \emph{choice}
which exists in the sequent calculus.
$$\begin{array}{lcll}
P,Q     &::=& \cdots\\
    &\mid& \Lcut{P}{`a}{x}{Q}\quad\textit{left-active cut}\\
    &\mid& \Rcut{P}{`a}{x}{Q}\quad\textit{right-active cut}
\end{array}
$$

\subsection{The computation}

There are $20$ reduction rules in $\x$ which  correspond to
cut-elimination in the sequent calculus and  which are split into
logical, activation and propagation clusters and not grouped
like~\cite{vBLL,vBL}. There are named to ease the comparison with $\astx$-rules.
\subsection*{Logical rules}
Logical rules say how to eliminate a cut.  They apply when the cut refers to two names which are
\emph{freshly introduced}.

\begin{definition}[Fresh introduction]~
\begin{itemize}
\item The term \ensuremath{P} freshly introduces \ensuremath{x} ~if~ \ensuremath{P=\Caps{x}{`a}} or \ensuremath{P=\Imp{Q}{`a}{x}{y}{R}},\\ with \ensuremath{x\notin N(Q), x\notin N(R)}.
\item The term \ensuremath{P} freshly introduces \ensuremath{`a} ~if~ \ensuremath{P=\Caps{x}{`a}} or
  \ensuremath{P=\Exp{x}{Q}{`b}{`a}},\\ with \ensuremath{`a\notin N(Q)}.
\end{itemize}

\end{definition}

Informally, names are freshly introducesd if they appear once and
only once, at the top level of their corresponding
terms.\footnote{This is more complex than in \ensuremath{\astx},
where the linearity condition guarantees that if a name occurs at
the top level, then it does not occur elsewhere.} The cut in this
position can not be activated. Logical rules are shown by
Figure~\ref{fig:xlogical}. \figurexlogical

\noindent The first two rules are \emph{renaming}. The last rule, called \emph{insertion},
defines an interaction between an importer and an exporter. It inserts an immediate subterm of an
exporter between two immediate subterms of an importer.
\subsection*{Activation rules}
Activation rules describe the non-determinism of classical cut-elimination.  If a cut refers to a
name which is not freshly introduced, one has to propagate it according to a chosen direction and
activation is then followed by propagation rules (see Figures \ref{fig:xmixL} and \ref{fig:xmixR}).
This choice has usually been bypassed in the previous interpretations, either by restricting the
reduction procedure (a very common one is to not allow cuts to pass over other active cuts), or by
giving priority to a specific strategy (like in~\cite{Danos}, by assigning colors to formulas).
Notice that the cut can be activated in one or the other direction when both conditions are
fulfilled at the same time, as shown by Figure \ref{fig:xactivation}.  This is a source of
non-confluence.%
\figurexactivation

\subsection*{Propagation rules}
\emph{Left} and \emph{right propagation} rules are given in Figures \ref{fig:xmixL} and
\ref{fig:xmixR}, respectively. These rules describe how a cut is pushed through a  term, but also address situations where deactivation, erasure and duplication occur. This means that in  $\x$, several actions can be defined by a single reduction rule. Take for example the rule from Figure~\ref{fig:xmixR} which involves propagation, duplication and deactivation:
\\[3mm]
$(\daggerR\mbox{-}prop\mbox{-}dupl\mbox{-}deact)\enskip\mbox{:}$\\\vspace*{-3mm}

\centerline{$\Rcut{P}{`a}{x}{(\Imp{Q}{`b}{x}{z}{R})}\rightarrow
    \Cut{P}{`a}{x}{(\Imp{(\Rcut{P}{`a}{x}{Q})}{`b}{x}{z}{(\Rcut{P}{`a}{x}{R})})}$}
\vspace*{3mm}
%
%
The rule labelled $(\daggerL\mbox{-}(c)\mbox{-}prop\mbox{-}deact)$
describes a subtle propagation over a capsule whose both names are
bound by cuts (nested cuts with an axiom). This rule is introduced
to prevent possible infinite reductions of syntactic
nature~\cite{DJS97,Urb00}. Rules $(\daggerL\mbox{-}gc)$ and
$(\daggerR\mbox{-}gc)$ collect garbage.

\figurexmixL
\figurexmixR

\subsection{The type system} The type assignment system for the~$\x$ calculus is given by Figure~\ref{fig:xtypes}.
\figurextypes

\subsection{Basic properties} It has been shown
in~\cite{BieUrb01} that the computation in $\x$ calculus can be
seen as proof-transformation (subject reduction property) and that
$\x$, although intrinsically non-deterministic and non confluent,
which are indeed properties of classical cut-elimination, is
strongly normalising.

\section{Erasure and duplication: the $\astx$ calculus}
\label{sec:astx}

This section presents the rules of untyped \mbox{$\astx$},
followed by the basic operational properties and the definition of
typing rules. Although it is presented here as a counterpart of
the implicative segment of the sequent calculus for classical
logic, the system can naturally be extended to encompass other
connectives as well~\cite{Zun07}.

\subsection{The syntax}
Names differ essentially from variables in~\mbox{$`l$-calculus}. The
difference lies in the fact that a variable can be substituted by an
arbitrary term, while a name can only be \emph{renamed} (that is,
substituted by another name). In $\astx$ the renaming is explicit,
which means that it is expressed within the language itself and is not
defined in the meta-theory. The reader will notice the presence of
hats on some names.  This notation has been borrowed from
\emph{Principia Mathematica} \cite{princ_mathem} and is used to denote
name binding.
%
%
\subsubsection*{Free names,  bound names and  $\astx$-terms}
\label{subsec:freenames}
Names can be free or bound. They are defined together with the set of
$\astx$-terms also called linear terms.

\medskip

\noindent\textbf{Linearity} \label{subsec:linearity}
In  $\astx$, we consider only \emph{linear} terms, which means:\\
-- Every name has at most one free occurrence, and \\
-- Every binder does bind an actual occurrence of a name (and thus
only one)

\begin{dfn}[Free Names and $\astx$- terms]\label{def:fn} %
The sets of \emph{free innames} and \emph{free
outnames} and the set of (well-formed) \emph{$\astx$-terms} are defined mutually recursively
in Figure~\ref{fig:freenames} and Figure~\ref{fig:linearterms}.
\end{dfn}
\figurefreenames %
\figurelinearterms %
By ``mutually recursive'' we mean that the definition of an $\astx$-term
needs the definition of the free names of its subterms and the
definition of the free names supposes that the structure of the
subterms is known.  We write $N(P)$ the \emph{set} of free names of
$P$, $ I(P)$ the \emph{sets} of free innames of~$P$, and $ O(P)$ the
set of free outnames of $P$. Thus \ensuremath{N(P)= I(P)\cup
  O(P)}.  A~name which occurs in $P$ and which
is not free is called a \emph{bound name}.  Notice that a construction
can bind two names. This can be either two innames, two outnames or an
inname and an outname.  Moreover, these names sometimes belong to
different subterms, as in the case of an importer or a cut. To denote
the binding of all names in one list, we use simply $\widehat{\cal
  I},~\widehat{\cal O}$.
It is sometimes needed to use ${\overline I}(P)$ instead of ${\cal
I}^P$, and similarly for outnames ${\overline O}(P)$, and names in
general ${\overline N}(P)$. The bar is used to denote that we see
the given set of names as a list, according to the total order
which can be defined for the set of names. To exclude a name from
a list, for instance, outname $\alpha$,  we write ${\cal
O}^{P}\backslash \alpha$.


\medskip

\noindent\textbf{Renaming} We define the operation
$\rename{P}{x}{y}$ which denotes the \emph{renaming} of a free
name $y$ in \ensuremath{P} by a fresh name $x$.  It is a
meta-operation which replaces a unique occurrence of a free name
by another free name. Therefore it is simpler than the
meta-substitution of \ensuremath{\lc}, which denotes the
substitution of a free variable (which can occur arbitrary number
of times) by an arbitrary term.


\medskip
\noindent \textbf{Indexing} We introduce a special kind of
renaming, called \emph{indexing}, in order to simplify the syntax
of the reduction rules. For example $P_i=ind(P,N(P),i)$ means that
$P_i$ is obtained by indexing free names in~$P$ by index $i$,
where $i\in N$. Simple notation $P_i$ for cases such as this one
will be used when possible. We assume that indexing always creates
fresh names. As we use it indexing preserves~linearity.


\medskip

\medskip
\noindent\textbf{Modules} A~module is a part of a term (not a
subterm) of the form
  $\Lcut{}{\alpha}{x}{Q}$ (\emph{left-module}) and $\Rcut{P}{\beta}{y}{}$ (\emph{right-module}) which  percolates through the structure of that term (and its subterms) during the computation, as specified by the so-called ``propagation rules''.  It resembles the explicit substitution.
We say that $`a$ and $y$ are the \emph{handles} of
$\Lcut{}{`a}{x}{Q}$ and $\Rcut{P}{`b}{y}{}$, respectively. Two
\emph{modules are independent} if the handle of one module does
not bind a free name inside the other module, and vice-versa, as
follows: \setlength{\arraycolsep}{2mm}
$$\begin{array}{c|c}
   \mbox{~~independent modules~~}       &   \mbox{~~conditions~~}\\[1mm]
   \hline\hline &\\[-2mm]
   \Lcut{}{`a}{x}{Q},~\Lcut{}{`b}{y}{R} &   \alpha\notin N(R),~\beta\notin N(Q)   \\[1mm]
   \Rcut{P}{`a}{x}{},~~\Rcut{Q}{`b}{y}{}&       x\notin N(Q),~y\notin N(P)      \\[1mm]
   \Rcut{P}{`a}{x}{},~~\Lcut{}{`b}{y}{R}&   x\notin N(R),~`b\notin N(P)
\end{array}$$

\medskip
\noindent\textbf{Convention on names} We adopt a convention on
names: ``a name is never both bound and free in the same term''.
Terms are defined up to \mbox{$`a$-conversion}, that is, the
renaming of bound names does not change them.

\medskip
Every  $\X$-term can be translated into an $\astx$-term, using
duplicators and erasers. For instance, $\Wr{\Caps{x}{`a}}{`a},$
which has two free occurrences of $`a$, can be represented in
$\astx$ by the term
$\Cr{\Wr{\Caps{x}{`a_1}}{`a_2}}{a_1}{`a_2}{`a}$ (notice the role
of a duplicator). The  term $\Exp{x}{\Caps{x}{\alpha}}{`b}{`g}$
binds no free name and corresponds to the $\astx$-term
$\Exp{x}{\Wr{(\Caps{x}{\alpha}}{`b})}{`b}{`g}$ (notice the role of
an eraser).

\begin{dfn}[Principal names] The following tables define te
so-called~\emph{principal names}.
\begin{center}
\begin{math}
\begin{footnotesize}
\setlength{\arraycolsep}{4mm}
    \begin{array}{c||c}
      \mbox{a term~}    & \mbox{L\mbox{-}princip. names}   \\[1mm] \hline\hline
      \Caps{x}{`a}          & x, `a      \\[1.5mm]
      \Exp{x}{P}{`b}{`a}    & `a      \\[1.5mm]
      \Med{P}{`a}{x}{y}{Q}  & x       \\[1.5mm]
      \Cut{P}{`a}{x}{Q}     & \textsf{none}
    \end{array}
  \qquad \qquad
  \begin{array}{c||c}
      \mbox{a term~}      & \mbox{S\mbox{-}princip. names}   \\[1mm] \hline\hline
      \Wl{P}{x}             & x            \\[1.5mm]
      \Wr{P}{`a}            & `a                \\[1.5mm]
      \Cl{P}{x_1}{x_2}{x}   & x     \\[1.5mm]
      \Cr{P}{`a_1}{`a_2}{`a}    & `a
    \end{array}
  \end{footnotesize}
\end{math}
\end{center}

We say that a name is \emph{principal} if it is either L-principal
(introduced by a logical term) or S-principal (introduced by a
structural term).
\end{dfn}

%



\begin{lem}
\label{lem:outname}
Every term has at least a free logical outname.
\end{lem}
\paragraph*{Proof:}The proof goes by routine induction on the structure of terms.\footnote{This would no longer be true if we were to extend the system with
negation, for details see~\cite{Zun07}.}\hfill$\boxempty$.

%
\begin{dfn}[Contexts]
\emph{Contexts} are formally defined as follows:
  \begin{center}
  \fbox{$
    \setlength{\arraycolsep}{2mm}
  \begin{array}{lclcl}
C\{\ \}&::=&       \{\ \}                    &\mid&  \Exp{x}{\{\ \}}{`b}{`a}      \\[1.4mm]
       &\mid&      \Med{\{\ \}}{`a}{x}{y}{Q} &\mid&  \Med{P}{`a}{x}{y}{\{\ \}}    \\[1.4mm]
       &\mid&      \Cut{\{\ \}}{`a}{x}{Q}    &\mid&  \Cut{P}{`a}{x}{\{\ \}}       \\[1.4mm]
       &\mid&      \Wl{\{\ \}}{x}            &\mid&  \Wr{\{\ \}}{`a}              \\[.4mm]
       &\mid&      \Cl{\{\ \}}{x}{y}{z}      &\mid&  \Cr{\{\ \}}{`a}{`b}{`g}      \\[1.4mm]
       &\mid&      C\{C\{\ \}\}
  \end{array}$}
\end{center}
\end{dfn}
\begin{rmk}
A context is a term with a hole in which another term can be placed. Therefore \ensuremath{\Cnt{P}} denotes placing the term \ensuremath{P} in the context \ensuremath{\Cnt{~}}.
\end{rmk}
\begin{rmk}We use \ensuremath{P=Q} to denote that the terms~\ensuremath{P} and~\ensuremath{Q} are syntactically~equal.
\end{rmk}

%
\begin{dfn}[Subterm relation \ensuremath{\preccurlyeq}]
\label{def:stm}
A term $Q$ is a \emph{subterm} of a term~$P$, denoted as $Q\preccurlyeq P$ if there is a context $\Cnt{~}$ such that $P=\Cnt{Q}$.
\end{dfn}
\begin{lem}The \emph{subterm} relation is reflexive, antisymmetric and transitive (i.e., is an order):
\label{lem:rat}
\begin{enumerate}
\item \emph{Reflexivity}\enskip $\stm{P}{P}$
\item \emph{Antisymmetry}\enskip If $\stm{P}{Q} \mbox{~and~} \stm{Q}{P} \mbox{~then~} P=Q$
\item \emph{Transitivity}:\enskip If $\stm{P}{Q} \mbox{~and~} \stm{Q}{R} \mbox{~then~} \stm{P}{R}$
\end{enumerate}
\end{lem}
\paragraph*{Proof:}
\begin{enumerate}
\item The first point is straightforward. If \ensuremath{\stm{P}{P}}, then by the subterm definition we have \ensuremath{\exists~\Cnt{~}} such that \ensuremath{P=\Cnt{P}}. This stands if we choose \ensuremath{\Cnt{~}} to be \ensuremath{\{~\}}.
\item Let \ensuremath{\stm{P}{Q}} and \ensuremath{\stm{Q}{R}}. By definition \ensuremath{\exists~\Cntp{~},\Cnts{~}} such that \mbox{\ensuremath{\Cntp{P}=Q}} and \ensuremath{\Cnts{Q}=P}. From \ensuremath{\Cntp{\Cnts{Q}}=Q} we derive \ensuremath{\Cntp{~}=\Cnts{~}=\{~\}}. Finally we can conclude \ensuremath{P=Q}.
\item On the one hand, from \ensuremath{\stm{P}{Q}} by definition we have: \ensuremath{\exists~\Cntp{~}} such that \ensuremath{\Cntp{P}=Q}.
On the other hand, from \ensuremath{\stm{Q}{R}} by definition we have: \ensuremath{\exists~\Cnts{~}} such that \ensuremath{\Cnts{Q}=R}. Thus,  \ensuremath{\Cnts{\Cntp{P}}=R} and therefore by definition we have \ensuremath{\stm{P}{R}}. \hfill\ensuremath{\boxempty}
\end{enumerate}


The following definition introduces the notion of a simple
context, i.e., a context which is not composed of other contexts.
Notice that it resembles the definition of context, except that
the cases~\ensuremath{\{~\}} and~\ensuremath{C\{C\{\ \}\}} are
omitted.
\begin{dfn}[Simple context]
A context $\Cnt{~}$ is said to be  \emph{simple} if $\Cnt{~}$ is one of the following:
  \begin{center}
  \fbox{$
    \setlength{\arraycolsep}{2mm}
    \begin{array}{lclcl}
C\{\ \}&:=&        \Exp{x}{\{\ \}}{`b}{`a}   &\mid&   \\[1.4mm]
       &\mid&      \Med{\{\ \}}{`a}{x}{y}{Q} &\mid&  \Med{P}{`a}{x}{y}{\{\ \}}    \\[1.4mm]
       &\mid&      \Cut{\{\ \}}{`a}{x}{Q}    &\mid&  \Cut{P}{`a}{x}{\{\ \}}       \\[1.4mm]
       &\mid&      \Wl{\{\ \}}{x}            &\mid&  \Wr{\{\ \}}{`a}              \\[.4mm]
       &\mid&      \Cl{\{\ \}}{x}{y}{z}      &\mid&  \Cr{\{\ \}}{`a}{`b}{`g}
  \end{array}$}
\end{center}
\end{dfn}
Using the definition of a simple context we formulate the notion of immediate subterm.
\begin{dfn}[Immediate subterm]
A term \emph{Q} is an \emph{immediate subterm} of \emph{P} if $P=\Cnt{Q}$ and $\Cnt{\ }$ is a simple context.
\end{dfn}

\begin{exa}
A term can have either one, two, or zero immediate subterms. For example, \ensuremath{\Med{Q}{`a}{x}{y}{R}} has two immediate subterms (these are \ensuremath{P} and \ensuremath{Q}),  \ensuremath{\Exp{x}{Q}{`b}{`a}} has one (a term~\ensuremath{P}), whilst \ensuremath{\Caps{x}{`a}} has zero immediate subterms.
\end{exa}
Using the definition of a context \ensuremath{\Cnt{~}}, we specify the notion of a \emph{context with two holes}.
\begin{dfn}[Context with two holes]~
  \begin{center}
  \fbox{$
    \setlength{\arraycolsep}{2mm}
    \begin{array}{lclcl}
C\{~,~\}&::=&        \Med{\{~\}}{`a}{x}{y}{\{~\}}   &\mid&  \Cut{\{\ \}}{`a}{x}{\{~\}} \\[1.4mm]
       &\mid&      C\{\Cnt{~},\Cnt{~}\} &\mid&  \Cnt{C\{~,~\}}
  \end{array}$}
\end{center}
\end{dfn}

\begin{dfn}[Simple context with two holes]~~
  \begin{center}
  \fbox{$
    \setlength{\arraycolsep}{2mm}
    \begin{array}{lclcl}
C\{~,~\}&::=&        \Med{\{~\}}{`a}{x}{y}{\{~\}}   &\mid&  \Cut{\{\ \}}{`a}{x}{\{~\}}
  \end{array}$}
\end{center}
\end{dfn}
Using this definition, the notion of \textit{immediate subterm}
can be naturally extended as to encompass the cases when we speak
about \textit{two immediate subterms}.

\begin{rmk} We allow the use of of $P=\CNT{P}{`a}{R}$to denote that
the term $P$ has $`a$ as principal name and $R$ as an immediate
subterm. Similarly for $P=\CNTD{P}{x}{R_1}{R_2}$
\end{rmk}

\begin{lem}The following holds:
\label{lem:ST}
\begin{enumerate}
\item If $`a\in N(P)$ then there exists a unique term $\stm{Q}{P}$ such that $`a$ is a principal name for $Q$.
\item If \ensuremath{x\in N(P)} then there exists a unique term \ensuremath{\stm{R}{P}} such that \ensuremath{x} is a principal name for \ensuremath{R}.
\end{enumerate}
\end{lem}
\begin{rmk} We will use the notation $Q^\alpha$ to specify that $Q$ has $\alpha$ as a principal name. Similarly, we use \ensuremath{R^x} to emphasize that \ensuremath{R} has \ensuremath{x} as a principal name.
\end{rmk}

\paragraph*{Proof:}We prove the first point. The proof goes by induction on the structure of a term \ensuremath{P} and case analysis.

Let \ensuremath{`a\in P}.
\begin{itemize}
\item Case:~ \ensuremath{`a} is a principal name for \ensuremath{P}. Then \ensuremath{Q=P}.
\item Case:~ \ensuremath{`a} is not a principal name for \ensuremath{P}. Then, either \ensuremath{P=\Cnt{R}} or \ensuremath{P=\Cnt{R_1,R_2}}, where \ensuremath{R,R_1} and \ensuremath{R_2} denote immediate subterms of~\ensuremath{P}.
    \begin{itemize}
    \item \ensuremath{P=\Cnt{R}}. By induction hypothesis, and since by linearity \ensuremath{`a\in R}, we have:
        \ensuremath{\exists\stm{Q}{R}} such that \ensuremath{`a} is a principal name for \ensuremath{Q}.
        By using transitivity (lemma~\ref{lem:rat}), from \ensuremath{\stm{Q}{R}} and \ensuremath{\stm{R}{P}}
        we infer \ensuremath{\stm{Q}{P}}.
    \item \ensuremath{P=\Cnt{R_1,R_2}}. By the linearity condition we know that \ensuremath{`a} belongs to either
        \ensuremath{N(R_1)} or \ensuremath{N(R_2)} (not to both). Thus we have two subcases, which correspond to the previous case.
        In the first case \ensuremath{\Cnt{R_1,R_2}} is seen as \ensuremath{\Cntp{R_1}}, where \ensuremath{\Cntp{~}=\Cnt{\{~\},R_2}},
        and in the second case as \ensuremath{\Cnts{R_2}}, where \ensuremath{\Cnts{~}=\Cnt{R_1,\{~\}}}.
        Recall that \ensuremath{R_1,R_2} are immediate subterms of \ensuremath{P} by definition.
    \end{itemize}
\end{itemize}
The second point of the lemma refers to innames instead of outnames, and the proof goes similarly.\hfill\ensuremath{\boxempty}

\paragraph{\bf Abbreviations} We introduce some abbreviations in order to represent
reduction rules in a convenient form.
\begin{center}
\begin{scriptsize}
    \begin{math}
      \begin{array}[c]{c@{\quad}c}
        \begin{array}[c]{c|c}
          \textrm{instead of } & \textrm{we write}\\\hline
          x_1{\odot}(...(\Wl{P}{x_n})...)
          & x_1\odot\ ...\Wl{P}{x_n} \\[2pt]
          (...(\Wr{P}{`a_1})\ ...){\odot}`a_n &
          \Wr{\Wr{P}{`a_1\ ...}}{`a_n}
        \end{array}
        &
        \begin{array}[c]{c|c}
          \textrm{instead of } & \textrm{we write}\\\hline
          \Cl{... \Cl{P}{y_n}{z_n}{x_n}...}
          {y_1}{z_1}{x_1}&
          \Clbindlistfin{P}{y_1}{y_n}{z_1}{z_n}{x_1}{x_n}    \\[4pt]
          \Cr{... \Cr{P}{`b_1}{`g_1}{`a_1}...}
          {`b_n}{`g_n}{`a_n} &
          \Crbindlistfin{P}{`b_1}{`b_n}{`g_1}{`g_n}{`a_1}{`a_n}
        \end{array}
      \end{array}
    \end{math}
\end{scriptsize}
\end{center}

\subsection{Reduction rules}
\label{sec:reductions} In this section we define the reduction
relation, $\rightarrow$. The set of reduction rules is rather large as it
captures classical cut-elimination.


Reduction rules are grouped into
\begin{itemize}
\item[1.] Activation rules (left and right)
\item[2.] Structural actions (left and right)
\item[3.] Deactivation rules (left and right)
\item[4.] Logical actions
\item[5.] Propagation rules (left and right)
\end{itemize}

\paragraph{Congruence rules} We assume some simple congruence rules which originate from the sequent calculus.

\noindent \begin{tabular*}{1.0\linewidth}{c@{\quad}c}
  \textit{Commuting names in a duplicator} & \textit{ Permuting independent duplicators} \\
          \fbox{$
        \begin{array}{ccc}
        \Cl{P}{x_1}{x_2}{x}            &\equiv&   \Cl{P}{x_2}{x_1}{x}\\
        \Cr{P}{`a_1}{`a_2}{`a}         &\equiv&   \Cr{P}{`a_2}{`a_1}{`a}
        \end{array}
    $}
& \fbox{$
\begin{array}{ccc}
  \Cl{\Cl{P}{y_1}{y_2}{y}}{x_1}{x_2}{x}
  &\equiv&
  \Cl{\Cl{P}{x_1}{x_2}{x}}{y_1}{y_2}{y}\\
  \Cr{\Cr{P}{`a_1}{`a_2}{`a}}{`b_1}{`b_2}{`b}   &\equiv&
  \Cr{\Cr{P}{`b_1}{`b_2}{`b}}{`a_1}{`a_2}{`a}\\
        \Cr{\Cl{P}{x_1}{x_2}{x}}{`a_1}{`a_2}{`a}          &\equiv&
            \Cl{\Cr{P}{`a_1}{`a_2}{`a}}{x_1}{x_2}{x}
\end{array}
$}
    \end{tabular*}

\noindent The conditions in the first rule treating the
duplicators are $y`;\{x_1,x_2\}$ and \mbox{$x`;\{y_1,y_2\}$} and
in the second $`b`;\{`a_1,`a_2\}$ and $`a`;\{`b_1,`b_2\}$. The
third rule allows us to drop parenthesis and use a simplified
notation
\begin{center}
  \(\Cboth{P}{x_1}{x_2}{x}{`a_1}{`a_2}{`a} \qquad \qquad \textrm{and
    more generally} \qquad \qquad \Cboth{P}{{\cal I}_1}{{\cal
      I}_2}{{\cal I}}{{\cal O}_1}{{\cal O}_2}{{\cal O}}\)
\end{center}
where $\mathcal{I}$ and $\mathcal{O}$ are lists of names. When
$~{\cal I}=(),~$ we write $~\Cr{P}{{\cal O}_1}{{\cal O}_2}{{\cal
O}}$. The case $~{\cal O}=()~$ is not possible as stated by
Lemma~\ref{lem:outname}.

\noindent \textit{When the names are triplicated, one can do it in
any order:}

\begin{center}\fbox{$
\begin{array}{ccc}
        \Cl{\Cl{P}{x_1}{x_2}{y}}{y}{x_3}{z}           &\equiv&
           \Cl{\Cl{P}{x_2}{x_3}{y}}{x_1}{y}{z}                          \\
        \Cr{\Cr{P}{`a_1}{`a_2}{`b}}{`b}{`a_3}{`g}     &\equiv&
           \Cr{\Cr{P}{`a_2}{`a_3}{`b}}{`a_1}{`b}{`g}
\end{array}
$}\end{center} This can be seen as an associativity of names bound
by a ternary duplicator.

\noindent \textit{Permuting the erasers:} The following rule
suggests that we may drop parenthesis and write $\
x\,\odot\,P\,\odot\,`a,$ and more generally we may write: ${\cal
I}\odot P\odot{\cal O}$.
\begin{center}\fbox{$
\begin{array}{ccc}
  y \odot x \odot P &`=& x \odot y \odot P  \\
  P \odot `a \odot `b &`=& P \odot `b \odot `a  \\
  \Wr{(\Wl{P}{x})}{`a}                            &`=&
  \Wl{(\Wr{P}{`a})}{x}
\end{array}
$}\end{center}

We now present the reduction rules of $\astx$ calculus.
%
\subsubsection*{1. Activation rules}
Activation rules hold the non-determinism of classical cut-elimination. More precisely, during the
process of cut-elimination sometimes we have to choose the left or the right subtree to push the cut
through.  This choice is captured by the activation rules, which require to extend the syntax with
new symbols called \emph{active cuts}.

\begin{dfn}[Active Cuts]
The syntax is extended with two \emph{active cuts}:
$$ P, Q ::= \ldots \mid \Lcut{P}{`a}{x}{Q} \mid  \Rcut{P}{`a}{x}{Q}$$
\end{dfn}



Activation rules are a potential source of non-confluence, an
intrinsic property of classical logic, as illustrated by
Example~\ref{ex:LAF}.

\begin{exa}
\label{ex:LAF}
Terms \ensuremath{\Lcut{P}{`a}{x}{Q}} and \ensuremath{\Rcut{P}{`a}{x}{Q}} are
essentially different. This becomes obvious in the example where both $`a$ and $x$ are introduced by erasers. Take
$$P= \Wr{M}{`a} \qquad\mbox{and}\qquad Q= \Wl{N}{x},$$ where $M$ and $N$ are arbitrary terms. Then we have:
$$
\begin{array}{lcl}
   \Lcut{(\Wr{M}{`a})}{`a}{x}{(\Wl{N}{x})} &\rightarrow & \Wboth{M}{{\cal I}^{N\setminus x}}{{\cal O}^N} \\
   \Rcut{(\Wr{M}{`a})}{`a}{x}{(\Wl{N}{x})} &\rightarrow & \Wboth{N}{{\cal I}^M}{{\cal O}^{M\setminus `a}}
\end{array}
$$
\end{exa}

\noindent This simple example is reminiscent of that of Lafont
\cite{GirardLafontTaylor89}.

\begin{rmk}
By constantly giving priority to either left or right activation,
we may remove the non-confluence from the calculus and thus obtain
two confluent subcalculi. In the case of $\lmu$-calculus, if one
gives priority to one of two sides, then one obtains a
call-by-name or a call-by-value calculus. In accordance to what was
noted for \ensuremath{\x} in~\cite{Urb01} - that
 this doesn't hold for $\x$, we suspect that it does not hold for $\astx$
either.
\end{rmk}
%
\subsubsection*{2. Structural actions}
\label{subs:str} Structural actions consist of four reduction
rules, specifying \emph{erasure} and \emph{duplication} by
referring to the situation when an active cut faces an eraser or a
duplicator. Structural actions are given in
Figure~\ref{fig:structuralactions}. These computational features
were studied extensively in the framework of intuitionistic logic
\cite{ReneDavid},\cite{KesLen07}.

\figurestructuralactions

Structural rules specifying duplication employ the so-called
simultaneous substitutions. Informally, simultaneous substitution
$\lsubs{`a_1}{`a_2}{x}{Q}$ in $P\lsubs{`a_1}{`a_2}{x}{Q}$ denotes
applying independent modules\footnote{These modules are
independent by definition of $\astx$-terms $\Lcut{}{`a_1}{x}{Q}$
and $\Lcut{}{`a_2}{x}{Q}$ on $P$ depending only on the occurrence
of $`a_1$ and $\alpha_2$ at the top level and the level of
immediate subterms of $P$. Similarly for $\rsubs{P}{`a}{x_1}{x_2}$
which is symmetrical.

\begin{dfn}[Simultaneous substitutions]
\label{def:ss}
We define simultaneous substitutions
$\lsubs{`a_1}{`a_2}{x}{Q}$ and $\rsubs{P}{`a}{x_1}{x_2}$ as
follows:
\begin{itemize}
\item \emph{Left simultaneous substitution}, $P\lsubs{`a_1}{`a_2}{x}{Q}$, is defined depending on the
structure of term $P$ to which it is applied:

  \begin{center}
  \fbox{$
    \setlength{\arraycolsep}{2mm}
  \begin{array}{l|l}
    P                        &  P\lsubs{`a_1}{`a_2}{x}{Q}\\
    \hline\\
    \CNT{P}{`a_1}{R}         &  \CBGenQ{\Cut{(\CNT{P}{`a_1}{\Lcut{R}{`a_2}{x_2}{Q_2}})}{`a_1}{x_1}{Q_1}}   \\[1mm]
    \CNT{P}{`a_2}{R}         &  \CBGenQ{\Cut{(\CNT{P}{`a_2}{\Lcut{R}{`a_1}{x_1}{Q_1}})}{`a_2}{x_2}{Q_2}}      \\[2mm]
    \CNT{P}{`b}{R}, `b\neq `a_1,`a_2
                             &
                             \CNT{P}{`b}{\Lcut{(\Cr{R}{`a_1}{`a_2}{`a})}{`a}{x}{Q}}\\[2mm]
    \CNTD{P}{}{R_1}{R_2}     &
                                \CBGenQ{\CNTD{P}{}{\Lcut{R_1}{`a_1}{x_1}{Q_1}}{\Lcut{R_2}{`a_2}{x_2}{Q_2}}},\\
                             &       \mbox{~if~}`a_1\in N(R_1),~`a_2\in N(R_2)   \\[1mm]
                             &  \CNTD{P}{} {\Lcut{(\Cr{R_1}{`a_1}{`a_2}{`a})}{`a}{x}{Q}}{R_2}
                                    \mbox{~if~}`a_1,a_2\in N(R_1)\\[1mm]
                             &  \CNTD{P}{} {R_1} {\Lcut{(\Cr{R_2}{`a_1}{`a_2}{`a})}{`a}{x}{Q}},
                                    \mbox{~if~}`a_1,a_2\in N(R_2)\\[1mm]
    \CNTD{P}{x}{R_1}{R_2}    & \mbox{Analogously to the previous case}
  \end{array}$}
\end{center}

Where $R,R_1,R_2$ denote immediate subterms of $P$, and where:
${\cal I}^Q =\overline I(Q)\setminus x,\enskip
                {\cal O}^Q =\overline O(Q) \mbox{~~and~~}
                Q_i = ind(Q,\  N(Q),\ i) \mbox{~~for~} i=1,2.$

\item \emph{Right simultaneous substitution}, $\rsubs{P}{`a}{x_1}{x_2}Q$, is defined
depending on the structure of term $Q$ to which it is applied:

  \begin{center}
  \fbox{$
    \setlength{\arraycolsep}{2mm}
  \begin{array}{l|l}
    Q                        &  \rsubs{P}{`a}{x_1}{x_2}Q\\
    \hline\\
    \CNT{Q}{}{R}             &  \CNT{Q}{}{\Rcut{P}{`a}{x}{(\Cl{R}{x_1}{x_2}{x})}}   \\[2mm]
    \CNTD{Q}{x_1}{R_1}{R_2}  &
    \CBGenP{\Cut{P_1}{`a_1}{x_1}{(\CNTD{Q}{x_1}{\Rcut{P_2}{`a_2}{x_2}{R_1}}{R_2})}},\\
                             &       \mbox{~if~}x_2\in N(R_1)   \\[1mm]
                             &
                             \CBGenP{\Cut{P_1}{`a_1}{x_1}{(\CNTD{Q}{x_1}{R_1}{\Rcut{P_2}{`a_2}{x_2}{R_2}})}},\\
                             &      \mbox{~if~}x_2\in N(R_2)   \\[1mm]
    \CNTD{Q}{x_2}{R_1}{R_2}  &  \mbox{Analogously to the previous case}   \\[2mm]
    \CNTD{Q}{}{R_1}{R_2}, y\neq x_1,x_2
                             &
                             \CBGenP{\CNTD{Q}{}{\Rcut{P_1}{`a_1}{x_1}{R_1}}{\Rcut{P_2}{`a_2}{x_2}{R_2}}},\\
                             &  \mbox{~if~} x_1\in N(R_1),~x_2\in N(R_2)\\
                             & \CNTD{Q}{}{\Rcut{P}{`a}{x}{(\Cl{R_1}{x_1}{x_2}{x})}}{R_2},
                             \mbox{~if~} x_1,x_2\in N(R_1)\\
                             & \CNTD{Q}{}{R_1}{\Rcut{P}{`a}{x}{(\Cl{R_2}{x_1}{x_2}{x})}},
                             \mbox{~if~} x_1,x_2\in N(R_2)\\[1mm]
    \CNTD{Q}{y}{R_1}{R_2}        &  \mbox{Analogously to the previous case}
  \end{array}$}
\end{center}
Where $R,R_1,R_2$ denote immediate subterms of $Q$, and  where:
${\cal I}^P =\overline I(P),\enskip {\cal O}^P =\overline
            O(P)\setminus `a \mbox{~~and~~}
            P_i = ind(P,\,N(P),\, i) \mbox{~~for~} i=1,2.$
\end{itemize}
\end{dfn}
}

%
\subsubsection*{3. Deactivation rules}
\label{sec:deact}

As we will see, active cuts will be blocked by L-principal names. Thus cuts must be deactivated to
continue to be distributed through the terms.  Deactivation rules are given in Figure
\ref{fig:deact}.
\figuredeact

Activation is dual of deactivation.  Activation and deactivation rules are designed is such that
they do not allow loops.  Indeed the side conditions do not allow an activation of a cut followed by a
deactivation of the same cut, or vice versa.

\subsubsection*{4. Logical actions}
The purpose of logical actions is to define reduction when L-principal names are involved in a cut. See Figure~\ref{fig:DLogicalRules}.
\figureDlogicalrules

The two first logical rules define the \emph{merge} of a capsule with another term using
  renaming $\ren{y}{x}$ which is a meta operation, which resembles the \ensuremath{\lambda}-calculus
  meta-substitution.  Renaming replaces simply a free name (unique by linearity) by another free
  name.  It does not essentially change the term.

The third logical action describes the direct interaction between
an exporter and an importer, which results in \emph{inserting} the
(immediate) subterm of an exporter between the two (immediate)
subterms of an importer.
%
\subsubsection*{5. Propagation rules}

Propagation rules describe the propagation of a cut through the structure
of a term. This is a step-by-step propagation (the reduction rules ``describe'' propagation).
It is important to note that  propagation of a cut over another inactive cut is possible, which allows an elegant representation of $\beta$-reduction.
The rules are divided into ``left'' and ``right'' symmetric groups,
see Figures~\ref{fig:LeftPropagation}~and~\ref{fig:RightPropagation}.

\figureleftpropagation
\figurerightpropagation

Observe for example the first rule in the left group. The rule is denoted as $(exp\daggerL-prop)$
and it shows how an active cut (in fact, a module $\Lcut{}{`b}{y}{R}$) enters from the right-hand
side through an exporter, up to its immediate subterm.  The rules which define propagation over an
exporter or a cut require side conditions to decide to which of the two immediate subterms the
module will go.

The rules which require additional explanations are
\ensuremath{(cut(c)\daggerL \mbox{-}prop)} and
\ensuremath{(\daggerR cut(c)\mbox{-}prop)}. These are the rules
which define an exception when performing propagation rules. They
handle the case of propagation over a cut with a capsule whose
both names are cut-names. If we exclude these rules from the
system, we could construct an infinite reduction sequence.

\begin{exa} An example of an infinite reduction sequence in absence of  \ensuremath{(cut(c)\daggerL \mbox{-}prop)} and \ensuremath{(\daggerR cut(c)\mbox{-}prop)} rules:
$$
\begin{array}{ll}
    &   \Cut{(\Cut{P}{`a}{x}{\Caps{x}{`b}})}{`b}{y}{R}\\
    \ra&    \Lcut{(\Cut{P}{`a}{x}{\Caps{x}{`b}})}{`b}{y}{R}\\
    \ra&    \Cut{P}{`a}{x}{(\Lcut{\Caps{x}{`b}}{`b}{y}{R})}\\
    \ra&    \Cut{P}{`a}{x}{(\Cut{\Caps{x}{`b}}{`b}{y}{R})}\\
    \ra&    \Rcut{P}{`a}{x}{(\Cut{\Caps{x}{`b}}{`b}{y}{R})}\\
    \ra&    \Cut{(\Rcut{P}{`a}{x}{\Caps{x}{`b}})}{`b}{y}{R}\\
    \ra&    \Cut{(\Cut{P}{`a}{x}{\Caps{x}{`b}})}{`b}{y}{R}
\end{array}
$$
\end{exa}

Besides that, the solution offered is intuitive as we would expect the terms
$$\Lcut{(\Cut{P}{`a}{x}{\Caps{x}{\beta}})}{\beta}{y}{R}\mbox{~~and~~}\Rcut{P}{`a}{x}{(\Cut{\Caps{x}{`b}}{`b}{y}{R})}$$
to reduce to the same term (which is in this case \ensuremath{\Cut{P}{`a}{y}{R} }).
%
%

\subsection{Operational properties}
The reduction system enjoys some desirable properties as expressed by the following lemma.

\begin{thm}[Basic properties of \ensuremath{\rightarrow}]~
\label{thm:xx}
 \begin{enumerate}
 \item Preservation of \emph{free names}:
            If~$P\rightarrow Q$ then  $\mbox{N}(P)=\mbox{N}(Q)$.
 \item Preservation of \emph{linearity}:
            If P is linear and $P\rightarrow Q$ then Q is linear.
 \end{enumerate}
\end{thm}


\paragraph*{Proof:}These properties can be confirmed by checking carefully each rule.\\
\mbox{~}\hfill$\boxempty$\vspace*{4mm}

\noindent Preservation of free names holds in $\astx$ due to the
use of erasers and duplicators in rewrite rules (like in
\mbox{$\lxr$}~\cite{KesLen07}). This property is sometimes
referred to as \emph{interface preservation} like in interaction
nets~\cite{Laf95}. The property of closure under reduction is a
minimal requirement, it is a kind of linearity preservation for
$\astx$-terms.

\paragraph*{Simplification rules}
We define the \emph{simplification rules}, denoted $\dashrightarrow$, which can be seen as an efficient way to simplify terms. They are not reduction rules as they do not involve cuts. The point is that applying a duplicator to an eraser is of no interest and can be
avoided by using simplification rules, as defined by:
 \begin{center}
\fbox{$
    \setlength\arraycolsep{2mm}
    \begin{array}{rcl}
            (s_L) &:& \Cl{\Wl{P}{z}}{y}{z}{x}\enskip~\dashrightarrow\enskip\rename{P}{x}{y}\\
            (s_R) &:& \Cr{\Wr{P}{\gamma}}{`b}{`g}{`a}\enskip\dashrightarrow\enskip\rename{P}{\alpha}{\beta}
    \end{array}
    $}\end{center}
They are run before reduction rules, that is, we give them higher
priority during computation. One can see them as a kind of garbage
collection, as they simplify computation by preventing the
situation when we duplicate a term to erase one or both copies in
the next step.  It is easy to see that the simplification rules
preserve free names, linearity and types. The rules can be given
in a more general way:
\begin{center}
\fbox{$
    \setlength\arraycolsep{2mm}
    \begin{array}{rcl}
            (s^g_L) &:& \Cl{\Wl{P}{{\I}_2}}{{\I}_1}{{\I}_2}{\I}\enskip~~\dashrightarrow\enskip\rename{P}{\I}{{\I}_1}\\
            (s^g_R) &:& \Cr{\Wr{P}{{\cal O}_2}}{{\cal O}_1}{{\cal O}_2}{\cal O}\enskip\dashrightarrow\enskip\rename{P}{{\cal O}}{{\cal O}_1}
    \end{array}
 $}\end{center}

\subsection{The type assignment system}
\label{sec:typedlanguage} %
We restrict now to terms to which we can attach types of the form:
\begin{eqnarray*}
  A,B ::= T \mid A \rightarrow B.
\end{eqnarray*}

The \emph{type assignment} of an $\astx$-term $P$ is expressed as $\witness{P}{`G}{`D}$,
  where $`G$ is the
antecedent whose domain  is made of free innames  of $P$ and $`D$
is the succedent whose domain is made of free outnames of $P$.
Contexts  are sets of pairs (name, formula). For example, $`G$ is
a set of type declarations for innames like~$x:A,~ y:B$,
while $`D$ as a set of declarations for outnames like
$\alpha:A$, $\beta : A\rightarrow B$, $\gamma : C$. Comma in the
expression \ensuremath{`G,`G'} stands for set union.

We will say that a term $P$ is \emph{typable} if there exist
contexts $`G$ and $`D$ such that $\witness{P}{`G}{`D}$ holds in
the system of inference rules given by Figure
\ref{fig:TypeSystem}. \figuretypesystem If we remove
term-decoration and names, we get the classical sequent system
$G1$ given in Figure~\ref{fig:sequentsystemone}.

\begin{exa}\label{ex:peirce}
An illustration could be the type assignment of the $\astx$-term which
codes the proof of Peirce's law.
\begin{center}
\begin{small}
\prooftree
\[
  \[
    \[
      \[
        \[\justifies
          \witness{\Caps{x}{`a_1}}{x:A}{`a_1:A}
          \using\emph{(ax)}
    \]
        \justifies
        \witness{\Wr{\Caps{x}{`a_1}}{`b}}{x:A}{`a_1:A,`b:B}
        \using \emph{(weak-R)}
      \]
      \justifies
      \witness{\Exp{x}{(\Wr{\Caps{x}{`a_1}}{`b})}{`b}{`g}}{}{`a_1:A,`g:A\ra B}
      \using \emph{($\ra\!$R)}
   \]
   \[ \justifies
      \witness{\Caps{y}{`a_2}}{y:A}{`a_2:A}
      \using \emph{(ax)}
   \]
   \justifies
   \witness{\Med{(\Exp{x}{(\Wr{\Caps{x}{`a_1}}{`b})}{`b}{`g})}{`g}{z}{y}{\Caps{y}{`a_2}}}{z:(A\ra B)\ra A}{`a_1:A,`a_2:A}
   \using \emph{($\ra\!$L)}
  \]
  \justifies
  \witness{\Cr{\Med{(\Exp{x}{(\Wr{\Caps{x}{`a_1}}{`b})}{`b}{`g})}{`g}{z}{y}{\Caps{y}{`a_2}}}{`a_1}{`a_2}{`a}}{z:(A\ra B)\ra A}{`a:A}
  \using \emph{(cont-R)}
\]
\justifies
\witness{\Exp{z}{(\Cr{\Med{(\Exp{x}{(\Wr{\Caps{x}{`a_1}}{`b})}{`b}{`g})}{`g}{z}{y}
{\Caps{y}{`a_2}}}{`a_1}{`a_2}{`a})}{`a}{`d}}{}{`d:((A\ra B)\ra A)\ra A}
\using \emph{($\ra\!$R)}
\endprooftree
\end{small}
\end{center}
\end{exa}\vspace*{8mm}
\begin{exa}
\label{ex:s}

The$\astx$-term which corresponds to $`lxyz.xz(yz)$, known as the
$S$-combinator of $\lc$, is the following:\footnote{Some parts of
terms are underlined to ease the reading.}

$$\Exp{\omega}{(\Exp{u}{(\Exp{x}{(\Cl{\underline{\Med{\Caps{x_2}{`e}}{`e}{w}{v}
    {\underline{(\Med{\underline{(\Med{\Caps{x_1}{`d}}{`d}{u}{y}{\Caps{y}{`b}})}}
    {`b}{v}{z}{\Caps{z}{`g}})}}}}{x_1}{x_2}{x})}{`g}{\eta})}{\eta}{\theta})}{\theta}{`a}$$

\end{exa}

\subsection*{The witness reduction property}
An $\astx$term is the interpretation of a proof in the sequent calculus.
If we use computations as proof-transformations, the property of witness reduction is essential.

\begin{thm}[Witness reduction]
\label{thm:subred}
Let $S$ be an $\astx$-term and ~$`G,`D$ two contexts. Then the following holds:
\begin{center}
    If\enskip $\witness{S}{`G}{`D}$\enskip and \enskip$S\rightarrow S'$,\enskip then \enskip $\witness{S'}{`G}{`D}$
\end{center}
\end{thm}
\begin{rmk}
 Linearity and free names are preserved (Theorem~\ref{thm:xx}).
\end{rmk}
\paragraph*{Proof:}
The proof is straightforward and goes by inspecting the reduction
rules, and by induction on the structure of terms~\cite{Zun07}.

\begin{thm}[\ensuremath{\dashrightarrow} preserves types] Simplification rules preserve types.
\label{thm:simp}
\begin{center}
    If\enskip $\witness{S}{`G}{`D}$\enskip and \enskip$S\dashrightarrow S'$,\enskip then \enskip $\witness{S'}{`G}{`D}$
\end{center}
\end{thm}
\paragraph*{Proof:}By analyzing the proof trees corresponding to \ensuremath{S} and \ensuremath{S'}, for both simplification rules.
\bigskip


\section{Explicit vs. implicit: relation between $\astx$ and $\x$}
\label{sec:xastx}

The $\astx$~calculus is a low-level language whose syntax is an extension of that of the~$\x$~calculus, and therefore its reduction steps decompose reduction steps of $\x$, which on its own is also a low level language.

The expressive power of $\x$ has been illustrated in \cite{vBLL}, by encoding various calculi, such as: $`l,~\lx$ and $\lambda\mu$. Also the~\mbox{$\x$~calculus} is encoded into $\lambda\mu$ in~\cite{AvB}. The first hint on how to relate $\lmm$ and Gentzen's sequent calculus for classical logic LK (which corresponds to $\x$) was already given by Curien and Herbelin in \cite{CurHer00}. It was studied in detail through \mbox{the~$\lambda\xi$-calculus} \cite{LenCBV}, where mutual embedings are presented. These results were used to give an elegant proof of strong normalization for the~$\lmm$-calculus.

%

Our view is that most of the features of the~$\x$~calculus can also be shown for $\astx$. Since the~$\astx$ calculus has a lower level of granularity, is expected to be at least as expressive as the $\x$ calculus. In case of potential implementation this model is better suited, since it introduces the possibility of controlling both duplication and erasure of parts of a program. In this chapter we study the relation between~$\astx$ and the following calculi; intuitionistic: $\lambda,~\lx$ and $\lxr$,  and classical: $\x$ and $\lmm$.
In this section we show the relation between $\x$-terms and $\astx$-terms. We present the encodings in both directions, and study the relation between the computations. It is shown that $\x$-reduction steps are decomposed into more atomic steps of $\astx$, due to the linearity and the presence of explicit terms for erasure and duplication. Finally, we study the relation between typing of $\x$-terms and typing of $\astx$-terms.

\subsection{From $\x$~to $\astx$}
We now describe how to encode $\x$-terms, which are possibly not
linear, into terms of the~$\astx$~calculus. Before doing that we
will introduce two operations to help us formulate the encoding.
They will be used in the formal definition and their only purpose
is to make definitions easier to read.
The first operation, denoted by $\circledcirc$, adds \emph{erasers} where needed.
\begin{dfn}[Potential eraser: $\circledcirc$]
\label{def:possw}
The operation $\circledcirc$ is defined as follows:
$$ x\circledcirc P \circledcirc `a\enskip =\enskip \left\{
               \begin{array}{ll}
                   \phantom{x\odot\ \,}P,       &\qquad  x,\alpha\in N(P) \\[1mm]
                   x\odot P,                &\qquad  x\notin N(P),\ `a\in N(P)\\[1mm]
           \phantom{x\odot\ \,}P\odot `a,       &\qquad  x\in N(P),\ `a\notin N(P)\\[1mm]
           x\odot P\odot `a,            &\qquad x,\alpha\notin N(P)
               \end{array}\right.
$$
\end{dfn}
The second operation, denoted by $\posscbothsmall{\ }{}{}$, adds
\emph{contractions} where needed. Typically this will happen when
encoding terms which have two immediate subterms, denoted
by~$C\{P,Q\}$, and it will be used to prevent the multiple
occurrences of names. This operation also improves the readability
of the encoding, although we could have used actual contractions.
\begin{dfn}[Potential contractions: $\posscbothsmall{\ }{}{}$]
\label{def:possc}
The operation $\posscbothsmall{\ }{}{}$ is defined as follows:
$$ \posscboth{C\{P,Q\}}{\mathcal I}{\mathcal O}\enskip =\enskip \left\{
               \begin{array}{ll}
                   C\{P,Q\},    &\qquad N(P)\cap N(Q)=\emptyset \\[1mm]
                   \Cboth{C\{P,Q\}}{{\mathcal I}_1}{{\mathcal I}_2}{\mathcal I}{{\mathcal O}_1}{{\mathcal O}_2}{\mathcal O}
            &\qquad \mbox{when~}  N(P)\cap N(Q)\neq\emptyset,\\
            &\qquad \mbox{where~} {\mathcal I}=I(P)\cap I(Q)\\
            &\phantom{\qquad\mbox{where}} {\mathcal O}=O(P)\cap O(Q)\\[1mm]
               \end{array}\right.
$$
\end{dfn}

\begin{dfn}
The encoding of $\X$-terms in $\astx$ is defined by induction, as presented by Figure~\ref{fig:encodingx}.
\end{dfn}
    \figureencodingx

\noindent Figure~\ref{fig:encodingx} defines the encoding of pure \ensuremath{\x}-terms in~\ensuremath{\astx}. Active cuts can be encoded in the following way:
$$  \encodeA{\Lcut{P}{`a}{x}{Q}}:= \Cbotho{\Lcut{(\Wro{\encodeA{P}}{`a})}{`a}{x}{(\Wlo{\encodeA{Q}}{x})}}{\mathcal I}{\mathcal O}$$
$$  \encodeA{\Rcut{P}{`a}{x}{Q}}:= \Cbotho{\Rcut{(\Wro{\encodeA{P}}{`a})}{`a}{x}{(\Wlo{\encodeA{Q}}{x})}}{\mathcal I}{\mathcal O}$$

\begin{rmk} Notice that if the $\x$-term is linear, i.e., if there is no need to use the operations $\circledcirc$ and $\posscbothsmall{\ }{}{}$, we get simply
    \begin{center}
    $\begin{array}{rcl}
    \encodex{\Caps{x}{\alpha}}  &=& \Caps{x}{\alpha}\\[2mm]
    \encodex{\Exp{x}{P}{`b}{`a}}    &=& \Exp{x}{\encodex{P}}{`b}{`a}\\[2mm]
    \encodex{\Imp{P}{`a}{x}{y}{Q}}  &=& \Imp{\encodex{P}}{`a}{x}{y}{\encodex{Q}}\\[2mm]
    \encodex{\Cut{P}{`a}{x}{Q}}     &=& \Cut{\encodex{P}}{`a}{x}{\encodex{Q}}
     \end{array}$
     \end{center}
\end{rmk}
\begin{rmk}
The encoding is defined in such a way that none of the free names is lost. Notice that this is not the case with \emph{occurrences} of free names. If a free name has multiple occurrences in $\x$-term, it will occur only once after the encoding.
\end{rmk}
\begin{lem}The encoding \ensuremath{\encodeA{~}} preserves the set of free names.
$$N(P)~=~N(\encodex{P})$$
\end{lem}
\begin{pf}By inspection of the encoding rules.\hfill$\boxempty$
\end{pf}

\begin{exa}Take for example $P=\Imp{(\Exp{x}{\Caps{x}{`a}}{`b}{`g})}{`g}{z}{y}{\Caps{y}{`a}}$, where $`a$ as a free name occurs twice, and \ensuremath{\widehat{`b}} does not bind an occurrence of a free name. The encoding gives:
$$\encodex{P}=\Cr{\Imp{(\Exp{x}{(\Wr{\Caps{x}{`a_1}}{`b})}{`b}{`g})}{`g}{z}{y}{\Caps{y}{`a_2}}}{`a_1}{`a_2}{`a}$$
where $`a$ has only one occurrence, and \ensuremath{\widehat{`b}} does bind an occurrence of a free name.
\end{exa}

\paragraph{Notation} We will sometimes annotate the arrow in
order to ease the reading: we use $\reduceastx$ to denote
$\astx$-reduction and $\reducex$ to denote $\x$-reduction.
Moreover $\redstar$ and $\redplus$ are used to denote zero or
more, and one or more reduction steps, respectively.

%
%

\paragraph*{Simulation of \ensuremath{\x}-reduction}
In what follows we show that the reduction rules of \ensuremath{\x}
can be simulated in~\ensuremath{\astx}. Initially we show that the notion of \emph{introduced name}
in~\ensuremath{\x} corresponds to the notion of \emph{L-principal name} in~\ensuremath{\astx}.

\begin{lem}
\label{lem:X}
The notion of \emph{introduced name} by a term in~\ensuremath{\x}, and that of \mbox{\emph{L-principal name}}
of a term in~\ensuremath{\astx}, correspond to each other.
\begin{enumerate}
\item If \ensuremath{`a} is freshly introduced by \ensuremath{S}, then \ensuremath{`a} is L-principal
for \ensuremath{\encodeA{S}}
\item If \ensuremath{`a} is L-principal for \ensuremath{S}, then \ensuremath{`a} is freshly introduced
by \ensuremath{\encodeB{S}}
\end{enumerate}
\end{lem}

\begin{pf}
\begin{enumerate}
\item
Case: \ensuremath{S=\Caps{x}{`a}}. We have \ensuremath{\encodeA{S}=\encodeA{\Caps{x}{`a}}=\Caps{x}{`a}}, and thus \ensuremath{`a} is L-principal for \ensuremath{\Caps{x}{`a}}.\\[2mm]
Case: \ensuremath{S=\Exp{x}{P}{`b}{`a}}. Since \ensuremath{`a} is freshly introduced \ensuremath{`a\notin N(P)}. We have  \ensuremath{\encodeB{S}=\encodeA{\Exp{y}{P}{`b}{`a}}}\ensuremath{=\Exp{y}{(\posswboth{\encodeA{P}}{y}{`b})}{`b}{`a}}, and thus by definition \ensuremath{`a} is L-principal for \ensuremath{\encodeA{S}}.
\item
Case: \ensuremath{S=\Caps{x}{`a}}. We have \ensuremath{\encodeB{S}=\encodeB{\Caps{x}{`a}}=\Caps{x}{`a}}, where \ensuremath{`a} is freshly introduced by  \ensuremath{\Caps{x}{`a}}, by definition.\\[2mm]
Case: \ensuremath{S=\Exp{x}{P}{`b}{`a}}. By linearity it stands that \ensuremath{`a\notin N(P)}. We have
\ensuremath{S=\encodeB{\Exp{y}{P}{`b}{`a}}}\ensuremath{=\Exp{y}{\encodeB{P}}{`b}{`a}}, and thus \ensuremath{`a} is freshly introduced by  \ensuremath{\encodeB{S}}.
\end{enumerate}
It is not difficult to check that the same holds for innames.\hfill\ensuremath{\boxempty}
\end{pf}


\begin{thm}[Simulation of $\x$-reduction]Let \ensuremath{P} and \ensuremath{P'} be \ensuremath{\x}-terms.
Then the following holds:
$$\mbox{If~} P ~\reducex~ P' \mbox{~~~then~~~} \encodex{P}~\reduceastxM~\Wboth{\encodex{P'}}
        {(\I^{P\setminus P'})}{({\mathcal O}^{P\setminus P'})}$$
\end{thm}
\begin{pf}
The proof goes by inspecting the reduction rules and
by induction on the structure of terms. We give the proof for some
reduction rules.

\paragraph*{Logical rules:}~\\[2mm]
\noindent$\bullet$\enskip Take the \ensuremath{(cap-ren)} rule:$~~\Cut{\Caps{y}{`a}}{`a}{x}{\Caps{x}{\beta}} \rightarrow \Caps{y}{\beta}$. We have:
\begin{center}
$
\begin{array}{lcl}
\encodeA{\Cut{\Caps{y}{`a}}{`a}{x}{\Caps{x}{\beta}}}
        &\triangleq& \Cut{\encodeA{\Caps{y}{`a}}}{`a}{x}{\encodeA{\Caps{x}{`b}}}\\
        &\triangleq& \Cut{\Caps{y}{`a}}{`a}{x}{\Caps{x}{`b}}    \\
        &\ra& \Caps{y}{b}\\
        &\triangleq& \encodeA{\Caps{y}{`b}}
\end{array}
$
\end{center}
\noindent$\bullet$\enskip Take the \ensuremath{(exp-ren)} rule: $~~\Cut{(\Exp{y}{P}{\beta}{`a})}{`a}{x}{\Caps{x}{\gamma}}
                    \rightarrow\Exp{y}{P}{\beta}{\gamma},\hfill `a\notin N(P).$ We have (assuming for simplicity that \ensuremath{`g\notin N(P)}):
\begin{center}
$
\begin{array}{lcl}
\encodeA{\Cut{(\Exp{y}{P}{\beta}{`a})}{`a}{x}{\Caps{x}{\gamma}} }
        &\triangleq& \Cut{\encodeA{\Exp{y}{P}{`b}{`a}}}{`a}{x}{\encodeA{\Caps{x}{`g}}}\\
        &\triangleq& \Cut{(\Exp{y}{(\pwb{\encodeA{P}}{y}{`b})}{`b}{`a})}{`a}{x}{\Caps{x}{`g}}\\
        &\ra& \Exp{y}{(\pwb{\encodeA{P}}{y}{`b})}{`b}{`g}\\
        &\triangleq& \encodeA{\Exp{y}{P}{`b}{`g}}
\end{array}
$
\end{center}
 \noindent$\bullet$\enskip Take the \ensuremath{(imp-ren)} rule: ~$\Cut{\Caps{y}{`a}}{`a}{x}{(\Med{Q}{\beta}{x}{z}{R})}
                    \rightarrow \Med{Q}{\beta}{y}{z}{R},\enskip\mbox{where}\\ x\notin N(Q),~ x\notin N(R).$ We have:
\begin{center}
$
\begin{array}{lcl}
    \encodeA{\Cut{\Caps{y}{`a}}{`a}{x}{(\Med{Q}{\beta}{x}{z}{R})}}
        &\triangleq& \Cut{\encodeA{\Caps{y}{`a}}}{`a}{x}{\encodeA{\Med{Q}{\beta}{x}{z}{R}}}\\
\multicolumn{3}{l}{\hspace*{2cm}\triangleq \Cut{\Caps{y}{`a}}{`a}{x}
                {(\posscboth{\Imp{(\posswr{\encodeA{Q}}{`b})}{`b}{x}{z}{(\posswl{\encodeA{R})}{z}}}{\mathcal I}{\mathcal O})}}\\
\multicolumn{3}{l}{\hspace*{2cm}\ra\posscboth{\Imp{(\posswr{\encodeA{Q}}{`b})}{`b}{y}{z}{(\posswl{\encodeA{R})}{z}}}{\mathcal I}{\mathcal O}}\\
\multicolumn{3}{l}{\hspace*{2cm}\triangleq\encodeA{\Imp{Q}{`b}{y}{z}{R}}}
\end{array}
$
\end{center}
For simplicity we assumed that \ensuremath{y\notin N(Q) \mbox{~and~} y\notin N(R)}.


\paragraph*{Activation rules}~\\[2mm]
\noindent$\bullet$\enskip Take the \ensuremath{(act-L)} rule: \ensuremath{~\Cut{P}{`a}{x}{Q} ~\rightarrow~ \Lcut{P}{`a}{x}{Q}}, \enskip if \ensuremath{`a} not freshly introduced by \ensuremath{P}. We have:
\begin{center}
$
\begin{array}{lcl}
\encodeA{\Cut{P}{`a}{x}{Q}}
        &\triangleq& \posscboth{\Cut{(\posswr{\encodeA{P}}{`a})}{`a}{x}{(\posswl{\encodeA{Q}}{x})}}{{\mathcal I}^{P\cap Q}}{{\mathcal O}^{P\cap Q}}\\[1mm]
        &\xrightarrow{Lem.\ref{lem:X}}& \posscboth{\Lcut{(\posswr{\encodeA{P}}{`a})}{`a}{x}{(\posswl{\encodeA{Q}}{x})}}{{\mathcal I}^{P\cap Q}}{{\mathcal O}^{P\cap Q}}\\[1mm]
        &\triangleq& \encodeA{\Lcut{P}{`a}{x}{Q}}
\end{array}
$
\end{center}
Similarly for the rule (\ensuremath{act-R}).

\paragraph*{Propagation rules}~\\[2mm]
\noindent$\bullet$\enskip Take the \ensuremath{(\daggerL-eras)} rule: \ensuremath{~\Lcut{\Caps{x}{`a}}{`b}{y}{R} ~\rightarrow~\Caps{x}{`a}, \mbox{~where~}  `a\neq `b}.\\
We will take into consideration the possibility that \ensuremath{x,`a\in N(R)}. Thus we have:
\begin{center}
$
\begin{array}{lcl}
\encodeA{\Lcut{\Caps{x}{`a}}{`b}{y}{R}}
        &\triangleq& \posscboth{\Lcut{(\Wr{\encodeA{\Caps{x}{`a}}}{`b})}{`b}{y}{(\posswl{\encodeA{R}}{y})}}{x}{`a}\\[1mm]
        &\ra& \posscboth{\Wboth{\encodeA{\Caps{x}{`a}}}{{\mathcal I}^R}{{\mathcal O}^R}}{x}{`a}\\[1mm]
        &\dashrightarrow& \Wboth{\encodeA{\Caps{x}{`a}}}{({\mathcal I}^{R}\setminus x)}{({\mathcal O}^R\setminus `a})
\end{array}
$
\end{center}

\medskip
\noindent$\bullet$\enskip Take the \ensuremath{(\daggerL-deact)} rule: \ensuremath{~\Lcut{\Caps{x}{`b}}{`b}{y}{R} ~\rightarrow~\Cut{\Caps{x}{`b}}{`b}{y}{R}}. We have:
\begin{center}
$
\begin{array}{lcl}
\encodeA{\Lcut{\Caps{x}{`b}}{`b}{y}{R}}
        &\triangleq& \poscl{\Lcut{\encodeA{\Caps{x}{`b}}}{`b}{y}{(\posswl{\encodeA{R}}{y})}}{x}\\[1mm]
        &\triangleq& \poscl{\Lcut{\Caps{x}{`b}}{`b}{y}{(\posswl{\encodeA{R}}{y})}}{x}\\[1mm]
        &\ra& \poscl{\Cut{\Caps{x}{`b}}{`b}{y}{(\posswl{\encodeA{R}}{y})}}{x}\\[1mm]
        &\triangleq& \encodeA{\Cut{\Caps{x}{`b}}{`b}{y}{R}}
\end{array}
$
\end{center}

\medskip
\noindent$\bullet$\enskip Take the \ensuremath{(\daggerL-prop)} rule: \ensuremath{~\Lcut{(\Exp{x}{P}{`g}{`a})}{`b}{y}{R}
        ~\rightarrow~\Exp{x}{(\Lcut{P}{`b}{y}{R})}{`g}{`a}}, \ensuremath{`a\neq `b}.
        We assume for simplicity \ensuremath{N(\Exp{x}{P}{`g}{`a})\cap N(R)=\emptyset}. We have:
\begin{center}
$
\begin{array}{lcl}
\encodeA{\Lcut{(\Exp{x}{P}{`g}{`a})}{`b}{y}{R}}
        &\triangleq& \Lcut{(\posswr{\encodeA{\Exp{x}{P}{`g}{`a}}}{`b})}{`b}{y}{(\posswl{\encodeA{R}}{y})}\\[1mm]
        &\triangleq& \Lcut{(\posswr{(\Exp{x}{(\posswboth{\encodeA{P}}{x}{`g})}{`g}{`a})}{`b})}{`b}{y}{(\posswl{\encodeA{R}}{y})}\\[1mm]
        &\equiv& \Lcut{(\Exp{x}{(\posswr{(\posswboth{\encodeA{P}}{x}{`g})}{`b})}{`g}{`a})}{`b}{y}{(\posswl{\encodeA{R}}{y})}\\[1mm]
        &\ra& \Exp{x}{(\Lcut{(\posswr{(\posswboth{\encodeA{P}}{x}{`g})}{`b})}{`b}{y}{(\posswl{\encodeA{R}}{y})})}{`g}{`a}\\[1mm]
        &\equiv& \Exp{x}{(\posswboth{(\Lcut{(\posswr{\encodeA{P}}{`b})}{`b}{y}{(\posswl{\encodeA{R}}{y})})}{x}{`g})}{`g}{`a} \\[1mm]
        &\triangleq& \Exp{x}{(\posswboth{(\Lcut{P}{`b}{y}{R})}{x}{`g})}{`g}{`a}\\[1mm]
        &\triangleq& \encodeA{\Exp{x}{(\Lcut{P}{`b}{y}{R})}{`g}{`a}}
\end{array}
$
\end{center}

\medskip
\noindent$\bullet$\enskip Take the
\ensuremath{(\daggerL\mbox{-}prop\mbox{-}dupl\mbox{-}deact)} rule:
\\ \ensuremath{~\Lcut{(\Exp{x}{P}{`g}{`b})}{`b}{y}{R}
        ~\rightarrow~\Cut{(\Exp{x}{(\Lcut{P}{`b}{y}{R})}{`g}{`b})}{`b}{y}{R}},
        and consider \ensuremath{\beta\in N(P)}. We assume for simplicity
        \ensuremath{N(P)\cap N(R)=\emptyset}, then we have:
\begin{center} $
\begin{array}{lcl}
    \encodeA{\Lcut{(\Exp{x}{P}{`g}{`b})}{`b}{y}{R}}
        &\triangleq& \Lcut{\encodeA{\Exp{x}{P}{`g}{`b}}}{`b}{y}{(\posswl{\encodeA{R}}{y})}\\[1mm]
    \multicolumn{3}{l}{\triangleq\Lcut{(\Cr{\underline{\Exp{x}{(\posswboth{(\encodeA{P}\ren{`b_1}{`b})}{x}{`g})}{`g}{`b_2}}}
            {`b_1}{`b_2}{`b})}
            {`b}{y}{(\posswl{\encodeA{R}}{y})}}\\[1.5mm]
    \multicolumn{3}{l}{\ra(\underline{\Exp{x}{(\posswboth{(\encodeA{P}\ren{`b_1}{`b})}{x}{`g})}{`g}{`b_2}})
            \lsubs{`b_1}{`b_2}{y}{(\posswl{\encodeA{R}}{y})}}\\[1mm]
    \multicolumn{3}{l}{\triangleq\posscboth{\Cut{(\Exp{x}{(\posswboth{(\Lcut{\encodeA{P}\ren{`b_1}{`b}}{`b_1}{y}
                {(\posswl{\encodeA{R}}{y})})}{x}{`g})}{`g}{`b_2})}{`b_2}{y}{(\posswl{\encodeA{R}}{y})}}
                {{\mathcal I}^R}{{\mathcal O}^R}}\\[1.5mm]
    \multicolumn{3}{l}{\triangleq \encodeA{\Cut{(\Exp{x}{(\Lcut{P}{`b}{y}{R})}{`g}{`b})}{`b}{y}{R}}}
\end{array}
$
\end{center}
%
\medskip
\noindent$\bullet$\enskip Take the
\ensuremath{(\daggerL\mbox{-}gc)} rule:
\ensuremath{\Lcut{P}{`a}{x}{Q} ~\ra~ P,\enskip\mbox{if}~`a \notin
N(P)}. Assume \ensuremath{N(P)\cap N(Q)=\emptyset}, we have:

\begin{center}
$
\begin{array}{lcl}
\encodeA{\Lcut{P}{`a}{x}{Q}}
        &\triangleq&
        \Lcut{(\Wr{\encodeA{P}}{`a})}{`a}{x}{(\posswl{\encodeA{Q}}{x})}\\[1mm]
        &\ra& \Wboth{\encodeA{P}}{\mathcal I^{\encodeA{Q}}}{\mathcal
        O^{\encodeA{Q}}},\\[1mm]
        && \mbox{which is what we expected.}
\end{array}
$
\end{center}
\medskip
Thus we are done with the proof.\ensuremath{\hfill\boxempty}

\end{pf}

\paragraph*{Preservation of types} We now show that the encoding preserves types.
In the typed $\x$~calculus contexts~$`G$ and~$`D$ may contain some
auxiliary pairs (name,type). This is due to the fact that
weakening is  implicit in \ensuremath{\X}, i.e., it is not
controlled explicitly. We have to keep that in mind when
formulating the lemma.

\begin{lem}[Preservation of types]
\label{lem:typepresx} If \ensuremath{P} is an arbitrary $\x$-term
such that $\witness{P}{`G}{`D}$, then
$$\witness{\Wboth{\encodex{P}}{((dom(\Gamma))\setminus I(P))}{((dom(\Delta))\setminus O(P))}}{`G}{`D}$$
\end{lem}
\begin{pf}
The proof works by case analysis and induction on
the structure of terms. We give the detail for encoding of capsule
and~exporter, whereas the other cases work the same way.

\medskip
\noindent$\bullet$\enskip Rule: \ensuremath{\encodeA{\Caps{x}{`a}}~:=~\Caps{x}{`a}}.\\
If \ensuremath{\Caps{x}{`a}\,\threedots `G\vdash `D} where \ensuremath{x:A\in `G} and \ensuremath{`a:A\in `D}, then, in \ensuremath{\astx} we have: \ensuremath{\Caps{x}{`a}\,\threedots x:A\vdash `a:A}, which is equivalent to:

\centerline{\ensuremath{\witness{\Wboth{\Caps{x}{`a}}{(dom(\Gamma)\setminus x)}{(dom(\Delta)\setminus `a)}}{`G}{`D}}}

\medskip
\noindent$\bullet$\enskip Rule: \ensuremath{\encodex{\Exp{x}{P}{`b}{`a}} ~:=~
                \poscr{\Exp{x}{(\posswboth{\encodex{P}}{x}{\beta})}{`b}{`a}}{`a}}.\\
If we assume the most generic case,
        namely for \ensuremath{x,`b\notin N(P)} and \ensuremath{\alpha\in N(P)}, then the encoding
        gives: \\
        \centerline{\ensuremath{\ensuremath{\encodex{\Exp{x}{P}{`b}{`a}}~:=~
        \Cr{\Exp{x}{(\Wboth{(\encodeA{P}\ren{`a_1}{`a})}{x}{`b})}{`b}{`a_1}} {`a_1}{`a_2}{`a}}}}

\medskip
On the one hand we have:
\begin{center}
\begin{small}
\[
\prooftree
    \witness{P}{`G}{\alpha:A\ra B, `D}
    \using (\ra R)\justifies
    \witness{\Exp{x}{P}{`b}{`a}}{`G}{`a:A\ra B,`D}
\endprooftree \]
\end{small}
\end{center}
where, as stated previously, \ensuremath{x:A\in `G,~`b:B\in `D}.

On the other hand,
\begin{center}
\begin{small}
\[
\prooftree
\[\[\[\[
    \witness{\encodeA{P}}{`G}{\alpha:A\ra B, `D}
    \using (ren)\justifies
    \witness{\encodeA{P}\ren{`a_1}{`a}}{`G}{`a_1:A\ra B,`D}
    \]
     \using (weak\mbox{-}L) \justifies
     \witness{\Wl{\encodeA{P}\ren{`a_1}{`a}}{x}}{`G,x:A}{`a_1:A\ra B, `D}
     \]
       \using (weak\mbox{-}R) \justifies
       \witness{\Wr{\Wl{\encodeA{P}\ren{`a_1}{`a}}{x}}{`b}}{`G,x:A}{`a_1:A\ra B, `b:B, `D}
       \]
        \using (\ra R) \justifies
        \witness{\Exp{x}{(\Wr{\Wl{\encodeA{P}\ren{`a_1}{`a}}{x}}{`b})}{`b}{`a}}{`G}{`a_1:A\ra B, `a_2:A\ra B,`D}
        \]
          \using (cont\mbox{-}R) \justifies
          \witness{\Cr{\Exp{x}{(\Wr{\Wl{\encodeA{P}\ren{`a_1}{`a}}{x}}{`b})}{`b}{`a}}{`a_1}{`a_2}{`a}}{`G}{`a:A\ra B, `D}
\endprooftree \]
\end{small}
\end{center}
~\hfill\ensuremath{\boxempty}
\end{pf}
%
%
%
%
%
%
%
\subsection{From $\astx$~to $\x$}
Now we investigate the opposite direction. We show how to represent \mbox{$\astx$-terms} by $\x$-terms and then we show how $\astx$-reductions are simulated by $\x$-reductions.
\begin{dfn}[Encoding $\astx$~into $\x$]
The encoding of $\astx$-terms in $\x$~calculus is defined inductively as shown by  Figure~\ensuremath{6.2}.
\end{dfn}
\figureencodingB
Encodings are defined without considering the active cuts but it is not difficult to extend it:
$$  \encodeB{\Lcut{P}{`a}{x}{Q}}:= \Lcut{\encodeB{P}}{`a}{x}{\encodeB{Q}}$$
$$  \encodeB{\Rcut{P}{`a}{x}{Q}}:= \Rcut{\encodeB{P}}{`a}{x}{\encodeB{Q}}$$

The encoding $\encodeB{~}$ does the opposite to $\encodex{~}$.
Namely, it simply removes erasers and duplicators from terms (some
renamings are also performed). That is the reason for a possible
decrease of free names after the encoding.

\begin{lem}[Properties of $\encodeB{~}$]
\label{lem:prop}
The encoding $\encodeB{~}$ satisfies the
following:
\begin{itemize}
\item[1.] $N(P)\subseteq N(\encodeB{P})$
\item[2.] $\encodeB{P}\ren{x}{y} = \encodeB{\rename{P}{x}{y}}$ if
$x\notin N(P)$
\end{itemize}
\end{lem}
\begin{pf} The former statement can be checked by carefully inspecting the encoding
rules, and the later by case analysis and induction on the
structure of terms.\hfill\ensuremath{\boxempty}
\end{pf}


The computation in $\astx$~is simulated by computation in $\x$ in
the way expressed by Theorem~\ref{thm:ast-x}; each reduction step
is mapped into one or more reduction steps.


%
%
%
%
%

\begin{lem}
\label{lem:names} Let $P$ be an $\astx$-term, and $\encodeB{P}$
its encoding in $\x$. Then the following holds:
\begin{itemize}
\item[1.] $`a,x\notin N(P) \ra `a,x\notin N(\encodeB{P})$
\item[2.] $`a,x\in N(\encodeB{P}) \ra `a,x\in N(P)$
\end{itemize}
\end{lem}

\begin{pf} Trivially by inspecting encoding rules. Names are lost
during encoding only if they are introduced in $\astx$ by
weakening.
\end{pf}


\begin{thm}[Simulating $\astx$-reduction]
\label{thm:ast-x} Let \ensuremath{P} and \ensuremath{P'} be
\ensuremath{\x}-terms.
Then the following holds:\\
 \centerline{If $P ~\reduceastx~ P'$ then $\encodeB{P}~\reducexplus~\encodeB{P'}$}
\end{thm}
\begin{pf}
The proof goes by inspecting the reduction rules and by induction
on the structure of terms. We provide the proof for several
reduction rules.

\paragraph*{Logical rules}~\\[2mm]
\noindent$\bullet$\enskip Take the \ensuremath{(ren-L)}
rule:$~~\Cut{\Caps{y}{`a}}{`a}{x}{Q} \rightarrow
\rename{Q}{y}{x}$.
We have:
\begin{center}
$
\begin{array}{lcl}
\encodeB{\Cut{\Caps{y}{`a}}{`a}{x}{Q}}
        &\triangleq& \Cut{\encodeB{\Caps{y}{`a}}}{`a}{x}{\encodeB{Q}}\\
        &\triangleq& \Cut{\Caps{y}{`a}}{`a}{x}{\encodeB{Q}}\\
        &\xrightarrow{\ensuremath{ren\mbox{-}L}}& \rename{\encodeB{Q}}{y}{x}\\
        &=& \encodeB{\rename{Q}{y}{x}}
\end{array}
$
\end{center}
\noindent$\bullet$\enskip Take the \ensuremath{(ren-R)} rule:
$~~\Cut{P}{`a}{x}{\Caps{x}{`b}}
                    \rightarrow \rename{P}{`b}{`a}$. We have :
\begin{center}
$
\begin{array}{lcl}
\encodeB{\Cut{P}{`a}{x}{\Caps{x}{`b}}}
        &\triangleq& \Cut{\encodeB{P}}{`a}{x}{\encodeB{\Caps{x}{`b}}}\\
        &\triangleq& \Cut{\encodeB{P}}{`a}{x}{\Caps{x}{`b}}\\
        &\xrightarrow{\ensuremath{ren\mbox{-}R}}& \rename{\encodeB{P}}{`b}{`a}\\
        &=& \encodeB{\rename{P}{`b}{`a}}
\end{array}
$
\end{center}

\paragraph*{Activation rules}~\\[2mm]
\noindent$\bullet$\enskip Take the \ensuremath{(act-L)} rule:
\ensuremath{~\Cut{P}{`a}{x}{Q} ~\rightarrow~ \Lcut{P}{`a}{x}{Q}},
\enskip if \ensuremath{`a} not L-principal for \ensuremath{P}. We
have:
\begin{center}
$
\begin{array}{lcl}
\encodeB{\Cut{P}{`a}{x}{Q}}
        &\triangleq& \Cut{\encodeB{P}}{`a}{x}{\encodeB{Q}}\\[1mm]
        &\xrightarrow{Lem.\ref{lem:X}}&  \Lcut{\encodeB{P}}{`a}{x}{\encodeB{Q}}\\[1mm]
        &\triangleq& \encodeB{\Lcut{P}{`a}{x}{Q}}
\end{array}
$
\end{center}
Similarly for the rule (\ensuremath{act-R}).

\paragraph*{Dectivation rules}~\\[2mm]
\noindent$\bullet$\enskip Take the
\ensuremath{(\daggerL\mbox{-}deact)} rule:
\ensuremath{~\Lcut{P}{`a}{x}{Q} ~\rightarrow~\Cut{P}{`a}{x}{Q}},
\enskip if \ensuremath{`a} is L-principal for \ensuremath{P}. We
have:
\begin{center}
$
\begin{array}{lcl}
\encodeB{\Lcut{P}{`a}{x}{Q}}
        &\triangleq& \Lcut{\encodeB{P}}{`a}{x}{\encodeB{Q}}\\
        &\xrightarrow{Lem.\ref{lem:X}}& \Cut{\encodeB{P}}{`a}{x}{\encodeB{Q}}\\
        &\triangleq& \encodeB{\Lcut{P}{`a}{x}{Q}}
\end{array}
$
\end{center}
Similarly for the rule (\ensuremath{\daggerR\mbox{-}deact}).

\paragraph*{Structural rules}~\\[2mm]
\noindent$\bullet$\enskip Take the
\ensuremath{(\daggerL\mbox{-}eras)} rule:
\ensuremath{~\Lcut{(\Wr{P}{`a})}{`a}{x}{Q} ~\rightarrow~
\Wboth{P}{{\cal I}^{Q}}{{\cal O}^{Q}}}. We have:
\begin{center}
$
\begin{array}{lcl}
\encodeB{\Lcut{(\Wr{P}{`a})}{`a}{x}{Q}}
        &\triangleq& \Lcut{\encodeB{\Wr{P}{`a}}}{`a}{x}{\encodeB{Q}}\\[1mm]
         &\triangleq& \Lcut{\encodeB{P}}{`a}{x}{\encodeB{Q}},~ `a\notin N(P)\\[1mm]
        &\xrightarrow{\ensuremath{\daggerL\mbox{-}gc}}&  \encodeB{P}\\[1mm]
        &\triangleq& \encodeB{\Wboth{P}{{\cal I}^{Q}}{{\cal O}^{Q}}}
\end{array}
$
\end{center}
{
\noindent$\bullet$\enskip Take the
\ensuremath{(\daggerL\mbox{-}dupl)} rule:
\ensuremath{~\Lcut{(\Cr{P}{`a_1}{`a_2}{`a})}{`a}{x}{Q}
~\rightarrow~ P\lsubs{`a_1}{`a_2}{x}{Q}}. We analyze here several
 cases of $P$.\\
- Take $P=\Exp{y}{R}{`g}{`b},~`b\neq `a_1,`a_2$. By definition of
$\astx$ terms $`a_1,`a_2\in N(R)$. Notice that $P$ is of the form
$\CNT{P}{`b}{R}$. We have:

\begin{center}
$
\begin{array}{lcl}
\encodeB{\Lcut{(\Cr{\Exp{y}{R}{`g}{`b}}{`a_1}{`a_2}{`a})}{`a}{x}{Q}}
        &\triangleq&
        \Lcut{\encodeB{\Cr{\Exp{y}{R}{`g}{`b}}{`a_1}{`a_2}{`a}}}{`a}{x}{\encodeB{Q}}\\
        &\triangleq&
        \Lcut{(\encodeB{\Exp{y}{R}{`g}{`b}}\ren{`a}{`a_1}\ren{`a}{`a_2})}{`a}{x}{\encodeB{Q}}\\
        &\triangleq&
        \Lcut{((\Exp{y}{\encodeB{R}}{`g}{`b})\ren{`a}{`a_1}\ren{`a}{`a_2})}{`a}{x}{\encodeB{Q}}\\
        &\triangleq&
        \Lcut{(\Exp{y}{(\encodeB{R}\ren{`a}{`a_1}\ren{`a}{`a_2})}{`g}{`b})}{`a}{x}{\encodeB{Q}}\\
        &\xrightarrow{\ensuremath{\daggerL\mbox{-}prop}}&
        \Exp{y}{(\Lcut{(\encodeB{R}\ren{`a}{`a_1}\ren{`a}{`a_2})}{`a}{x}{Q})}{`g}{`b}\\
        &\triangleq&
        \Exp{y}{(\Lcut{\encodeB{\Cr{R}{`a_1}{`a_2}{`a}}}{`a}{x}{Q})}{`g}{`b}\\
        &\triangleq&
        \Exp{y}{\encodeB{\Lcut{(\Cr{R}{`a_1}{`a_2}{`a})}{`a}{x}{Q}}}{`g}{`b}\\
        &\triangleq&
        \encodeB{\Exp{y}{(\Lcut{(\Cr{R}{`a_1}{`a_2}{`a})}{`a}{x}{Q})}{`g}{`b}}\\
        &\triangleq&
        \encodeB{\CNT{P}{\beta}{R}\lsubs{`a_1}{`a_2}{x}{Q}},\mbox{~when~}`b\neq
        `a_1,`a_2,\\
        && \mbox{by def. of simultaneous subst. on page~\pageref{def:ss}.}
\end{array}
$
\end{center}

\noindent - Take $P=\Exp{y}{R}{`g}{`a_1}$. By definition of
$\astx$ terms $a_2\in N(R), `a_1\notin N(R)$. Notice that $P$ is
of the form $\CNT{P}{`a_1}{R},~`a_2\in R$. We have:

\begin{center}
$
\begin{array}{lcl}
&~&
\encodeB{\Lcut{(\Cr{\Exp{y}{R}{`g}{`a_1}}{`a_1}{`a_2}{`a})}{`a}{x}{Q}}\\
        &\triangleq&
        \Lcut{\encodeB{\Cr{\Exp{y}{R}{`g}{`a_1}}{`a_1}{`a_2}{`a}}}{`a}{x}{\encodeB{Q}}\\
        &\triangleq&
        \Lcut{(\encodeB{\Exp{y}{R}{`g}{`a_1}}\ren{`a}{`a_1}\ren{`a}{`a_2})}{`a}{x}{\encodeB{Q}}\\
        &\triangleq&
        \Lcut{((\Exp{y}{\encodeB{R}}{`g}{`a_1})\ren{`a}{`a_1}\ren{`a}{`a_2})}{`a}{x}{\encodeB{Q}}\\
        &\triangleq&
        \Lcut{(\Exp{y}{(\encodeB{R}\ren{`a}{`a_2})}{`g}{`a})}{`a}{x}{\encodeB{Q}}\\
        &\xrightarrow{\ensuremath{\daggerL\mbox{-}prop\mbox{-}dupl\mbox{-}deact}}&
        \Cut{(\Exp{y}{(\Lcut{(\encodeB{R}\ren{`a}{`a_2})}{`a}{x}{\encodeB{Q}})}{`g}{`a})}{`a}{x}{\encodeB{Q}}\\
        &=&
        \Cut{(\Exp{y}{(\Lcut{\encodeB{R\ren{`a}{`a_2}}}{`a}{x}{\encodeB{Q}})}{`g}{`a})}{`a}{x}{\encodeB{Q}}\\
        &\triangleq&
        \Cut{(\Exp{y}{\encodeB{\Lcut{R\ren{`a}{`a_2}}{`a}{x}{Q}}}{`g}{`a})}{`a}{x}{\encodeB{Q}}\\
        &\triangleq&
        \Cut{\encodeB{\Exp{y}{(\Lcut{R\ren{`a}{`a_2}}{`a}{x}{Q})}{`g}{`a}}}{`a}{x}{\encodeB{Q}}\\
        &\triangleq&
        \encodeB{\CBGenQ{\Cut{(\Exp{y}{(\Lcut{R}{`a_2}{x_2}{Q_2})}{`g}{`a_1})}{`a_1}{x_1}{Q_1}}}\\
        &\triangleq&
        \encodeB{\CNT{P}{\alpha_1}{R}\lsubs{`a_1}{`a_2}{x}{Q}},\\
        && \mbox{by def. of simultaneous subst. on page~\pageref{def:ss}.}
\end{array}
$
\end{center}

\noindent - Take $P=\Imp{R_1}{`g}{y}{z}{R_2}$, and assume
$`a_1,`a_2\in N(R_1)$. Notice that $P$ is of the form
$\CNTD{P}{y}{R_1}{R_2}$. We have:

\begin{center}
$
\begin{array}{lcl}
        &~&
        \encodeB{\Lcut{(\Cr{\Imp{R_1}{`g}{y}{z}{R_2}}{`a_1}{`a_2}{`a})}{`a}{x}{Q}}\\
        &\triangleq&
        \Lcut{\encodeB{\Cr{\Imp{R_1}{`g}{y}{z}{R_2}}{`a_1}{`a_2}{`a}}}{`a}{x}{\encodeB{Q}}\\
        &\triangleq&
        \Lcut{(\encodeB{\Imp{R_1}{`g}{y}{z}{R_2}}\ren{`a}{`a_1}\ren{`a}{`a_2})}{`a}{x}{\encodeB{Q}}\\
        &\triangleq&
        \Lcut{((\Imp{\encodeB{R_1}}{`g}{y}{z}{\encodeB{R_2}})\ren{`a}{`a_1}\ren{`a}{`a_2})}{`a}{x}{\encodeB{Q}}\\
        &\triangleq&
        \Lcut{(\Imp{(\encodeB{R_1}\ren{`a}{`a_1}\ren{`a}{`a_1})}{`g}{y}{z}{\encodeB{R_2}})}{`a}{x}{\encodeB{Q}}\\
        &\triangleq&
        \Lcut{(\Imp{\encodeB{\Cr{R_1}{`a_1}{`a_2}{`a}}}{`g}{y}{z}{\encodeB{R_2}})}{`a}{x}{\encodeB{Q}}\\
        &\xrightarrow{\ensuremath{\daggerL\mbox{-}prop\mbox{-}dupl_1}}&
        \Imp{(\Lcut{\encodeB{\Cr{R_1}{`a_1}{`a_2}{`a}}}{`a}{x}{\encodeB{Q}})}{`g}{y}{z}{(\Lcut{\encodeB{R_2}}{`a}{x}{\encodeB{Q}})}\\
        &\xrightarrow{\ensuremath{\daggerL\mbox{-}gc}}&
        \Imp{(\Lcut{\encodeB{\Cr{R_1}{`a_1}{`a_2}{`a}}}{`a}{x}{\encodeB{Q}})}{`g}{y}{z}{\encodeB{R_2}}\\
        &\triangleq&
        \Imp{\encodeB{\Lcut{(\Cr{R_1}{`a_1}{`a_2}{`a})}{`a}{x}{Q}}}{`g}{y}{z}{\encodeB{R_2}}\\
        &\triangleq&
        \encodeB{\Imp{(\Lcut{(\Cr{R_1}{`a_1}{`a_2}{`a})}{`a}{x}{Q})}{`g}{y}{z}{R_2}}\\
        &\triangleq&
        \encodeB{\CNTD{P}{y}{R_1}{R_2}\lsubs{`a_1}{`a_2}{x}{Q}},
                \mbox{~when~}`a_1,`a_2\in N(R_1)
\end{array}
$
\end{center}

\noindent - Take $P=\Imp{R_1}{`g}{y}{z}{R_2}$, and assume $`a_1\in
N(R_1),~`a_2\in N(R_2)$.
We have:

\begin{center}
$
\begin{array}{lcl}
        &~&
        \encodeB{\Lcut{(\Cr{\Imp{R_1}{`g}{y}{z}{R_2}}{`a_1}{`a_2}{`a})}{`a}{x}{Q}}\\
        &\triangleq&
        \Lcut{\encodeB{\Cr{\Imp{R_1}{`g}{y}{z}{R_2}}{`a_1}{`a_2}{`a}}}{`a}{x}{\encodeB{Q}}\\
        &\triangleq&
        \Lcut{(\encodeB{\Imp{R_1}{`g}{y}{z}{R_2}}\ren{`a}{`a_1}\ren{`a}{`a_2})}{`a}{x}{\encodeB{Q}}\\
        &\triangleq&
        \Lcut{((\Imp{\encodeB{R_1}}{`g}{y}{z}{\encodeB{R_2}})\ren{`a}{`a_1}\ren{`a}{`a_2})}{`a}{x}{\encodeB{Q}}\\
        &\triangleq&
        \Lcut{(\Imp{(\encodeB{R_1}\ren{`a}{`a_1})}{`g}{y}{z}{(\encodeB{R_2}\ren{`a}{`a_2})})}{`a}{x}{\encodeB{Q}}\\
        &\xrightarrow{\ensuremath{\daggerL\mbox{-}prop\mbox{-}dupl_1}}&
        \Imp{(\Lcut{\encodeB{R_1}\ren{`a}{`a_1}}{`a}{x}{\encodeB{Q}})}{`g}{y}{z}{(\Lcut{\encodeB{R_2}\ren{`a}{`a_2}}{`a}{x}{\encodeB{Q}})}\\
        &\triangleq&
        \Imp{\encodeB{\Lcut{R_1}{`a_1}{x}{Q}}}{`g}{y}{z}{\encodeB{\Lcut{R_2}{`a_2}{x}{Q}}}\\
        &\triangleq&
        \encodeB{\CBGenQ{\Imp{(\Lcut{R_1}{`a_1}{x_1}{Q_1})}{`g}{y}{z}{(\Lcut{R_2}{`a_2}{x_2}{Q_2})}}}\\
        &\triangleq&
        \encodeB{\CNTD{P}{y}{R_1}{R_2}\lsubs{`a_1}{`a_2}{x}{Q}},
               \mbox{~when~}`a_1\in N(R_1),`a_2\in N(R_2)
\end{array}
$
\end{center}

The proof for propagation group of rules is
straightforward.$\hfill\boxempty$

}
\end{pf}

\subsection{Strong normalisation of $\astx$}

Exploiting the strong normalisation property of simply typed $\x$ \cite{Urb01}, we prove that $\astx$  is strongly normalising. We first prove that the previously defined encoding of $\astx$ into $\x$ preserves typeability.

\begin{lem}[Preservation of types]
\label{lem:types} For an arbitrary $\astx$-term $P$ such that
$\witness{P}{`G}{`D}$, it stands
$$\witness{\encodeB{P}}{\Gamma}{\Delta}$$
\end{lem}
\begin{pf}
By induction on typing derivations along the lines of
Lemma~\ref{lem:typepresx}.\hfill$\boxempty$
\end{pf}

This section presents the proof of strong normalisation for
$\astx$ calculus.

\begin{thm}[Strong Normalisation]
\label{thm:SN} The reduction system of $\astx$ is strongly
normalising on simply-typed terms.
\end{thm}

\begin{pf}
Let  $\witness{P}{`G}{`D}$. Assume that $P$ is not strongly normalising, which means that there is an infinite reduction starting with $P$

$$P  ~\reduceastx~ P_1 ~\reduceastx~  \ldots ~\reduceastx~ P_n ~\reduceastx~ \ldots$$
then by Theorem~\ref{thm:ast-x},
$$\encodeB{P}~\reducexplus~\encodeB{P_1} \reducexplus~ \ldots \reducexplus~\encodeB{P_n} \reducexplus~ \ldots $$
On the other hand according to Lemma~\ref{lem:types},
$$ \witness{\encodeB{P}}{\Gamma}{\Delta}$$
and the fact that $\x$ calculus is strongly normalising on typed
terms (\cite{Urb01}), we conclude that $\encodeB{P}$ is strongly normalising, which contradicts the assumption. Hence, $P$ is strongly normalising.
\hfill$\boxempty$
\end{pf}

\nocite{Zun07,LesZun08WRS,fromXtoPi,vanBakel12,KesRen11,KesRen09}


\section{Conclusions}
\label{sec:conclusion}

  We have presented two calculi implementing the Curry-Howard correspondence for classical logic
  sequent calculi.  The first one, called $\x$ provides terms for sequent proofs in the calculus
  $G3$ and a description of cut elimination by reductions.  A type system for this calculus assigns
  types to terms. The type of a term is the proposition that the proof associated with the term
  proves.  We designed the calculus $\astx$ in some sense as an extension of $\x$ with rules for explicit
  structural rules known in the sequent calculus $G1$ as \emph{weakening} and \emph{contraction}.  In $\astx$, the
  operator associated with weakening is an erasure  and the operator associated with contraction is a duplication.  
  Like $\x$, $\astx$ is associated with a type system to represent proofs in a sequent
  calculus with weakening and contraction.  We have explored the connection between the
  logic calculus $G3$ (resp. $G1$) and its implementation $\x$ (resp. $\astx$).  We have also shown
  how $\x$ can be embedded in $\astx$ and vice-versa.  
  As a low level language, it reveals details in both, structure of
  terms and computation, but in the same time this explicitness yields
  the essence of classical proofs and classical computations.  We know
  that the $`l$-calculus is the framework of functional sequential
  programming and $\astx$ can be seen as an extension of
  $`l$-calculus. An interesting direction for future work could be to
  explore the connections between $\astx$ and non deterministic
  distributed calculi like what has been done by van Bakel,
  Cardelli and Vigliotti~\cite{fromXtoPi}.


\bibliographystyle{elsarticle/elsarticle-harv}

\end{document}